# The Wage Premium of Communist Party Membership: Evidence from China[☆]

PLAMEN NIKOLOV[a,b,†]    HONGJIAN WANG[a]    KEVIN ACKER[c]

[☆]We thank the editor for comments that significantly improved the paper. Plamen Nikolov gratefully acknowledges research support by The Harvard Institute for Quantitative Social Science, the Economics Department at the State University of New York (Binghamton), the Research Foundation for SUNY at Binghamton. We thank Matthew Bonci for outstanding research support. All remaining errors are our own.

[†]Corresponding Author: Plamen Nikolov, Department of Economics, State University of New York (Binghamton), Department of Economics, 4400 Vestal Parkway East, Binghamton, NY 13902, USA. Email: pnikolov@binghamton.edu

[a] State University of New York (Binghamton)
[b] Harvard Institute for Quantitative Social Science
[c]The Johns Hopkins University School of Advanced International Studies and The Hopkins-Nanjing Center

# The Wage Premium of Communist Party Membership: Evidence from China

Social status and political connections could confer large economic benefits to an individual. Previous studies focused on China examine the relationship between Communist party membership and earnings and find a positive correlation. However, this correlation may be partly or totally spurious, thereby generating upwards-biased estimates of the importance of political party membership. Using data from three surveys spanning more than three decades, we estimate the causal effect of Chinese party membership on monthly earnings in in China. We find that, on average, membership in the Communist party of China increases monthly earnings and we find evidence that the wage premium has grown in recent years. We explore for potential mechanisms and we find suggestive evidence that improvements in one's social network, acquisition of job-related qualifications and improvement in one's social rank and life satisfaction likely play an important role. (*JEL* D31, J31, P2)

# I. Introduction

Because party membership can confer large economic benefits, numerous economics studies have examined the how political and social status influence later-life economic outcomes. Recent empirical research in developing countries documents the causal impact of political status on investments (Fishman, 2001), firm value (Erkal & Kali, 2012), and wages (Li et al., 2007). China's rapid economic development and its one-party government system make for an ideal context to examine the interplay between political status and economic outcomes. Previous studies document that political status in China can provide numerous economic benefits (Morduch & Sicular, 2000, Li et al., 2007, Appleton et al., 2003, Wu, Wu & Rui, 2010). Party membership provides access to a social network that can improve employment outcomes. Previous economics studies have examined how a social network or improved social status can influence one's economic outcomes.[1]

In this study, we estimate the wage premium associated with membership in the Chinese Communist Party over the span of almost four decades. We use data from the China Household Income Project (CHIP), China Housing Survey (CHS), and the Chinese General Social Survey (CGSS). The Communist party of China is the largest political party in the world (Yuen, 2013). With over eighty million members and growth of an average of one million members per year, the Chinese Communist party will likely remain the largest party in the world (Yuen, 2013).[2] Previous estimates of the effect of political status on economic outcomes in China rely on ordinary least squares (OLS) estimations using observational data. Because such study designs are not fully equipped to disentangle the influence of party membership from the influence of other background characteristics on employment outcomes, they do not detect true causal effects of party membership on wages. To estimate the effect of Communist party membership on monthly earnings, we employ a propensity score matching method (Rosenbaum & Rubin 1983, 1984; Imbens 2004; Imbens & Wooldridge 2009; Imbens & Rubin 2014; Abadie *et al.* 2003;

---

[1] Montgomery (1991) models the role of social networks in employee referrals for better employment prospects and how networks could improve the employer-employee match. Other studies document the effect of various types of social network on employment outcomes – e.g., fraternity and sorority membership at an American university (Marmaros & Sacerdote, 2001), ethnic networks (Patacchini & Zenou, 2012), and "*guanxi*" networks in China on job allocation (Bian, 1994).

[2] More than 80 percent of upperclassmen at Chinese universities apply for Communist party membership (The Economist, 2014). Instead of joining the party for its ideals, many aspire to join it because of their beliefs that party membership will provide better job market prospects – party membership is increasingly seen as a resume booster (Yuen, 2013).

Angrist & Krueger 2000; Angrist & Pischke 2009). We also perform various robustness checks to bolster the credibility of our results. Although the matching methods offers several important advantages over the ordinary least squares (OLS) method, it is also prone to bias in the presence of selection on observables. To gauge the presence of potential bias based on unobservables , we bound the magnitude of the potential bias using Rosenbaum bounds.

We report three main findings. First, using a propensity score matching technique, we find that Communist party members, on average, earn approximately a little over 20 percent more in monthly earnings compared to non-members. Our estimated effect size adds to existing empirical literature using data from developing countries that documents substantial wage benefits premium associated with political status and social connections (Siddique, 2010; Madheswaram and Attewell, 2007; Das and Dutta, 2007). We further bolster the credibility of our estimated effect sizes with several additional robustness checks: we examine the sensitivity of our results with respect to the matching algorithm method, the estimation technique and finally we gauge the potential selection bias due selection of unobservable variables using a Rosenbaum bounds method. Second, because we rely on data over three decades and three major data surveys, we find suggestive evidence that this wage premium has grown modestly over the last three decades. In both the ordinary least squares and propensity score matching results, we detect evidence that the wage premium associated with party membership has increased. Finally, we explore for the relative importance of various channels in explaining the linkage between party membership and better wage outcomes. Party membership could translate into better employment outcomes – on the intensive and extensive margin -- because of several distinct mechanisms.  Party membership could confer access to a wider social network of local party members and members could get a higher number of job referrals or access to better types of jobs (Morduch & Sicular, 2000). These job referrals could serve as a device that assists in a better employer-employee match in the long-run. Finally, some employment opportunities in provincial and local governments are only open to party members (The Economist, 2014). Based on the available longitudinal data from one of the data surveys, we explore for four main channels: strength of one's social network, human capital acquisition as one becomes a party member, improvement in one's social rank and overall life satisfaction. We provide suggestive evidence that at least three important channels likely exert very strong positive influence on one's wages when one joins the Communist Party: we see strong evidence that the strongest impact on wage seems to be through an improvement the likelihood of one holding a government job, improvement in one's position within one's job hierarchy, and improvement in one's overall social rank. We also see some evidence (though not statistically significant) that members report being happier than non-members. In summary, all of the findings provide robust evidence that political connections can play an important economic role in the world's most populous economy.

Our study contributes the empirical literature on the economic benefits of political affiliation in at least four major ways.[3] First, we examine the progression the earnings premium of party membership over almost four decades in China. Previous studies on the topic rely exclusively on single datasets and therefore can isolate the magnitude of the wage premium due party membership only for a single year. We combine data from three major Chinese data surveys – the China Household Income Project (CHIP), China Housing Survey (CHS), and the Chinese General Social Survey (CGSS) – that collect data on party membership, one earnings and contain rich information on various labour market factors. Because we rely on multiple datasets over more than three decades, we can shed light on the relationship over a longer period of time. Second, we provide convincing empirical evidence that the earnings premium associated with party membership has either remained constant or has grown. Third, we show that when we estimate the earnings premium using the propensity matching method[4], we find evidence that the OLS estimates are consistently lower than the estimated effect size using the matching method. This finding is consistent with positive self-selection (Roy, 1951) into party membership.[5] Finally, and perhaps the most novel aspect of this study, is that we are able to shed light on important channels that underlie the relationship between one's party affiliation and earnings. We detect suggestive evidence that when one joins the party, one's growing social network, higher perceived social and job status and access to better jobs likely plays a very large role in driving higher earnings.

---

[3] Extensive research on the caste system in India as status differentiation documents persistent economic impacts (Singh, 2010, Das & Dutta, 2007, Attewell & Madheswaran, 2007, Siddique, 2010, Eswaran et al., 2013). Siddique (2010) finds that, on average, low-caste applicants have to send 20 percent more resumes to get the same amount of call-backs. Madheswaram and Attewell (2007) and Das and Dutta (2007) document difference between low-caste and high-caste workers from 15 percent to 30 percent. Previous studies also examine the effect of status in the context of organizations and social networks (Patacchini & Zenou, 2012, Marmaros and Sacerdote, 2001). Patacchini and Zenou (2012) find that living in an area with a high concentration of those with the same ethnic status as you increase the probability of finding a job through social contacts. Marmaros and Sacerdote (2001) find that social networks play a large role in finding a job for students at Dartmouth College. Closely tied with the idea of social networks is social capital, defined by the density and diversity of one's social networks (Growiec & Growiec, 2015). Zhang and Anderson (2014) find that "bridging" social capital yields positive economic returns, while "bonding" social capital does not[3]; Growiec and Growiec (2015) find that the returns to "bridging" social capital are inverted U-shaped.

[4] We estimate the causal effect of Communist party membership on wages by using a propensity score matching technique (Rosenbaum & Rubin, 1983, 1984; Imbens, 2004; Imbens & Wooldridge, 2009; Imbens & Rubin, 2014; Abadie et al., 2003; Angrist & Krueger, 2000; Angrist & Pischke, 2009).

[5] *Positive* selection implies that workers that party members generally have better unobservable characteristics that drive selection into both party membership and higher earnings.

The remainder of this paper is structured as follows. Section II provided background on the party membership process, the various channels underpinning the relationship between party affiliation and higher earnings, and previous empirical studies. Section III details the data. Section IV outlines our identification strategy and Section V discusses results. Section VI discusses various robustness checks. Section VII concludes.

## II. Background

### A. Communist Party in China

The Communist Party of China is the largest political party in the world. With a membership of almost 88 million, the number of people in the communist party is greater than the population of Germany. However, with a total population of 1.37 billion, the party consists of less than 7 percent of the population (South China Morning Post, 2015). Recently, the party has been trending towards a more youthful and better educated member base. In 2014, 2.1 million new members were accepted. About 81 percent of new members were under 35, and 39 percent had university degrees, which is 2.6 percentage points more than in 2013. The Communist Party of China is overwhelmingly male, with only 21.7 million female members out of a total of 88 million, about 25 percent. As for the occupations of the members, 26 million (30 percent) identify as working in agriculture, 7.3 million (8 percent) identify as workers, 12.5 million (14 percent) identify as professionals, 9 million (10 percent) identify as administrative staff, and 7.4 million (8 percent) identify as government workers. This leaves 25.8 million (29 percent) that do not identify with any of the above categories (South China Morning Post 2015).

The Communist party of China's mission is to be a "vanguard party," which entails being comprised of the most capable and ideologically enlightened few of society (Xia, 2006). Towards that objective, the party admission process is long, rigorous, and selective. In 2014 the acceptance rate for party membership was the same as that of the Ivy League. Only 2 million were admitted to the party out of an applicant pool of 22 million (McMorrow, 2015). Selection for membership in the communist party is based on a broad range of criteria, including family background, academic performance, and party loyalty (Bian et al., 2001).

The party initiation process goes through several steps. The first step for a party hopeful is to compose a letter to the party organization affiliated with one's school or work. In this letter, one must make a case for their membership. Successful letters include descriptions of academic success, offices held, and involvement in party-related activities. The letter should also portray one's loyalty to the communist party and knowledge of party history. If, based on this letter, one is picked by the party organization to become an applicant, or "activist" according to party rhetoric, training begins (McMorrow, 2015). Successful activists frequently participate in political activities, including engaging in community service and attending lectures given by the local branch secretary. Activists are required to report regular "self-assessments," and are also assessed by liaisons assigned by the party organization (Bian et al., 2001). Activists are assessed on their political loyalty, work or academic performance, social activities and personal relationships (Li et al., 2007). In addition, activists must attend a party class led by a professor of political thought. The curriculum includes the party's history, structure and political purpose (McMorrow, 2015). This process continues for approximately 2 to 3 years (Li et al., 2007). Once the party determines the time has come for a final decision to be made regarding a particular activist's transition into full membership, the final assessments take place. The activist must take a two-hour written examination on Marxism and Chinese Communist ideology, including that of Mao Zedong and Deng Xiaoping. In addition, the local party organization interviews the activist's peers and superiors to gain more insight into the activist's personal quality and political character (McMorrow, 2015). A panel of party members also interviews the activist directly, inquiring about their political activities and quizzing them on their knowledge of recent party statements and events (McMorrow, 2015). This final decision making process culminates in a closed-door meeting where the local party organization tasks its members with judging political performance, personal history, and family background (Bian et al., 2001). This meeting ends in a vote on the activist's admission into the party. If the party organization votes in favour of the applicant's membership, that applicant is registered as a tentative party member, with full membership granted after a year of probation (Bian et al., 2001).[6]

---

[6] During the probationary period, members participate in all party meetings and activities, but cannot vote on party initiatives or be considered for party positions (Bian et al., 2001). These party activities include meetings to study party documents or discuss national policies (McMorrow, 2015). During the probationary period, members are closely monitored by the party organization. However, if a year passes and the member has not broken any party rules or engaged in any subversive activities, they become full members (Li et al., 2007).

Once admitted, one's membership in the Communist Party of China acts as a foot in the door to becoming one of China's administrative elite. Government positions and positions in state-run organizations with political or managerial authority are only open to party members. Access to positions at a certain level of hierarchy is controlled by authorities at the next higher level. Personnel offices of local party committees keep dossiers on all party members under their jurisdiction, recording successes and failures, and which are then consulted when examining a candidate for a position (Bian et al., 2001).

**B.  Conceptual Framework**

First, it could be the case that once individuals become party members they learn useful skills via exposure to other members or learn valuable skills directly from other party members.[7] This channel is akin formal education being a form human capital: the distinction is that skill acquisition occur through other party members as opposed to formal schooling as it occurs in the *human capital* theory (Mincer, 1974; Willis, 1986).[8]

Second, party membership increases one's social capital by providing access to a social network that can yield valuable connections. These connections can then lead to referrals for jobs. The connections one gets through party membership can be considered a type of "bridging" social capital, which Zhang and Anderson (2014) determined leads to economic returns. Bian (1994) finds that party members are more likely to use social connections to find jobs than non-members. Additionally, party membership yields higher status connections than those found outside the party, which leads to referrals for higher status jobs (Bian, 1994). The importance of social networks in labour markets is pervasive and well documented. Granovetter (1973, 1995) found in a survey of residents of a Massachusetts town that over 50 percent of jobs were obtained through social contacts. Earlier work by Rees (1966) found numbers of over 60 percent in a similar study. Exploration in a large number of studies documents similar figures for a variety of occupations, skill levels, and socioeconomic backgrounds.

Third, party membership can translate into higher wages simply because of the fixed cost to obtaining certain jobs -- some high paying jobs are only open to party members. Many employment opportunities in provincial and local governments are only open to party members (The Economist, 2014). In addition, all manager level jobs or higher in state-run organizations are only open to party members (Bian, 1994). Although our data does not enable us to disentangle the importance of each of these channels, all of these channels translate membership directly into better employment prospects either through the extensive margin (the likelihood of landing a job), through a better job match that results in higher wages or in higher wages. We return to providing suggestive evidence on the importance of each of these channels later.

---

[7] The rigorous process of applying to and joining the party confers valuable transferrable skills in the Chinese labour market (Pan, 2010).
[8] Membership in the Communist party can also serve as a signal of one's unobserved ability (Spence, 1974). Signals are important in hiring because the process of matching unemployed workers to suitable jobs with vacancies, and the bargaining process by which salaries are set are hindered by imperfect information (Yashiv, 2007). It is practically impossible for employers to assess how good a match might turn out to be because in addition to ability, job-specific productivity is also driven by personalities, fit with job culture, and other employee-specific preferences (Velasco, 2011). Signals like party membership can therefore, help employers to make hiring decisions. The Communist party attempts to attract the brightest and best of Chinese society, employers can take party membership as a signal for higher ability, which can lead to higher paying jobs. However, our survey sources do not provide direct information on one's cognitive abilities.

### C. Towards More Causal Estimates of the Wage Premium of Party Membership

Our paper builds on an empirical literature that attempts to measure the effect of party membership on earnings. One early study in this literature is Morduch & Sicular (2000), a study which attempts to estimate the effect of being a Communist party member on household income in rural China using an OLS approach.[9] Morduch & Sicular (2000) find that households with a cadre member exhibit approximately 20 percent higher earnings. However, they also fail to detect differences in household income between households with just one party member versus households with no party members. These results seem to suggest that monetary benefit to being a party member are conferred and mediated only through higher levels of Communist party involvement. Because of the study's OLS-based research design, it cannot claim true causal effects of party membership.[10]

Li et al. (2007) uses data from the Chinese Twins Survey, which relies on twins in five cities in China. The study examines the effect of party membership on income within pairs of twins where one is a member and the other is not a member. They use twins to account for observable and unobservable differences that result in omitted variable bias in OLS-based observational data estimation. Their study estimates that the income of party members was 10 percent higher than non-party members, but when accounting for "within-twin-pair fixed effects" to control for differences in ability and family background, the study detects no difference in income between party members and non-party members.

---

[9] The study measured involvement in the CPC on two levels: 1) The household includes a party member and 2) the household includes a party cadre. A party cadre is someone who "holds an official position of political or administrative leadership" (Morduch & Sicular, 2000). Morduch & Sicular (2000) argue that in order for transition in a socialist country like China to succeed, rank-and-file officials (in the case of China, party cadres) need to have some incentive to administrate the changes of the transition even if it could have a negative effect on their political and economic status. Subsequently, one should expect to see positive household income effects on a household having a cadre member.

[10] Morduch & Sicular (2000) only examine rural China, where the networks and credentials that Communist party membership confers may have very little benefit because of a lack of employment opportunities. In addition, because they rely on OLS design, their estimation is likely plagued by omitted variable bias due party membership being related to ability or other time-variant individual specific characteristics that influence earnings. In contrast, in this paper we examine the effect of party membership in two of China's major cities, Tianjin and Shanghai, where the benefits conferred by social and party networks and credentials are substantially higher.

Although twin models present some advantages (as highlighted above), they are also plagued with some important limitations as acknowledged by Griliches (1979) and Neumark (1999). For example, Neumark (1999) extends the analysis in Griliches (1979), and shows that the within-twin IV estimator amplifies the bias from any omitted ability differences between twins, relative to the standard within-twin estimator. Moreover, the paper clearly shows that if omitted ability biases cross-section estimates of the return to schooling upward, and is not fully removed by differencing within twin pairs, then the within-twin IV estimator is upward biased (possibly substantially) relative to the standard within-twin estimator, and possibly also relative to the cross-section estimator. This point is relevant for any application in which instrumental variables estimation is used for differenced data, when the differencing may not fully eliminate the omitted variable. Neumark (1999) remarks that the rationale for within-twin estimation of the return to schooling is the presumption that identical twins have equal ability, which drops out of the within-twin difference. However, this does not explain the source of schooling differences within twin pairs.

The notion that within-twin estimates provide a "natural experiment" for estimating the return to schooling is based on the assumption that schooling differences within twin pairs represent random (true) variation. However, once alternative reasons for schooling differences among twins are considered (and if twins are identical, we must wonder about the source of schooling differences between them), the conditions for this experiment to be valid may be violated, and may, in some circumstances, imply that the bias in within-twin estimates is greater than that in cross-sectional estimates.

The second important limitation of twin studies rests on an important assumption which some empirical paper contradict. For example, in the related empirical literature on the returns to schooling, twins-based estimates of the return to schooling have featured prominently. Their unbiasedness hinges critically on the assumption that within-pair variation in schooling is explained by factors unrelated to wage earning ability. Sandewall, Cesarini and Johanneson (2014) develops a framework for testing this assumption and shows, in a large sample of monozygotic twins, that the twins-based estimated return to schooling falls if adolescent IQ test scores are included in the wage equation. Using birth weight as an alternative proxy for ability yields qualitatively similar results. The results of Sandewall, Cesarini and Johanneson (2014) thus cast strong doubts on the validity of twins-based estimates.[11] The counterpart assumption in our context is that the unbiasedness in twin studies on the effect of communist party membership on earnings hinges critically on the assumption that within-pair variation in party membership is explained by factors unrelated to wage earning ability. Twin studies on the effect of party membership on earnings could be flawed because there is likely non-random variation in party membership correlated with ability differences of the twins.

## III. Data and Survey Sources

### A. Survey Data

We draw on data from three major Chinese surveys that contain information on party membership, earnings and various labour market factors: the China Housing Survey, the Chinese Household and Income Project survey and the Chinese General Social Survey.

***The China Housing Survey (CHS).*** First, we use data from the China Housing Survey (CHS) carried out in 1993 in Tianjin and Shanghai. This cross-sectional dataset was collected at the household level, and 2,096 households were interviewed in total. The neighbourhoods from which the households were drawn and the addresses of the households themselves were randomly selected. In addition, the survey was completed by randomly selecting and interviewing a respondent from within the household. The surveys were almost identical in each city, and were conducted simultaneously. In both cities, the surveys were authorized by the government, and the response rates were close to 100 percent. However, the sampling methods differ slightly for Tianjin and Shanghai (Bian et al. 1999).[12,13] The information collected of primary concern to this study was the wealth of demographic information, including ethnicity, gender, marriage status, education and neighbourhood.[14]

---

[11] In the context of educational attainment and estimating the returns to schooling, twin study designs have two strong downsides: (1) they can exacerbate measurement error (Light and Flores-Lagunes 2006; Ashenfelter and Krueger 1994, Behrman et al. 1994, Miller et al. 1995), and (2) Non-random variation in schooling correlated with ability differences of the twins – see Sandewall, Cesarini and Johannesson (2014). Both of these issues are very much potential threats to the validity of empirical estimates of party membership on earnings based on twin study designs.

***The Chinese Household and Income Project Survey (CHIP).*** Our second survey source is the Chinese Household Income Project (CHIP) a survey conducted between 1988 and 2013 of about 8,000 rural households (representing some 35,000 individuals) and almost 7,000 urban households (approximately 22,000 members). To track the dynamics of income distribution in China, the CHIP conducted five waves of household surveys, in 1989, 1996, 2003, 2008 and 2013 covering income and expenditure information.[15] In this study, we use data from 1988 and 2002. These surveys were carried out as part of a collaborative research project on incomes and inequality in China organized by Chinese and international researchers, with assistance from the National Bureau of Statistics (NBS).

The urban survey covered 10,000 households containing 29,262 individuals selected from 302 cities in sixteen provinces, whereas the rural survey covered 13,000 households containing 51,847 individuals selected from 287 counties in sixteen provinces. The migrant survey covered nearly 5,000 households containing 8,404 individuals selected from fifteen cities in nine provinces. To obtain a nationally representative sample, the provinces were selected from four distinct regions to reflect variations in economic development and geography.[16][17]

---

[12] In Tianjin, data collection was coordinated by the Tianjin academy of Social Science. Households from a randomly selected set of 125 neighbourhoods were interviewed. One neighbourhood was chosen from each sub district of the city. The addresses for the households chosen were randomly selected from the Tianjin household registration system. The Tianjin surveys were incorporated into an annual Tianjin municipal government survey called the "One Thousand Household Survey." In total, 1,042 households were surveyed in Tianjin. The sample from Tianjin is slightly biased towards male heads of households, but distributions on other characteristics are close to those of the general population in Tianjin as reported in the census (Bian et al., 1998).

[13] In Shanghai, data collection was coordinated by the Shanghai Academy of Social Science. Like in Tianjin, one neighbourhood from every sub district of the city was randomly selected, for a total of 110 neighbourhoods. The households interviewed were drawn from these neighbourhoods, and their addresses were randomly selected from the Shanghai census. In total, 1,054 households were surveyed in Shanghai. The distributions of the sample's characteristics are close to those of the general population of Shanghai (Bian et al., 1998).

[14] Intended primarily as a housing survey, CHS elicited information on length of stay and frequency of moves, physical style of housing and organization of housing space, accessibility of utilities, amount of rent/payment and work unit subsidies, strategies for obtaining better housing, and neighbourhood support networks. Other items covered include job opportunity, collective welfare programs, employee training programs, and relationships with others in the work unit and with the work unit leader. Information was collected on up to 9 of the respondent's household members, as well as the respondent's spouse, parents and in-laws, regardless of whether they lived in the household.

[15] The CHIP survey was conducted in 1988, 1995, 2002, 2007 and 2013 respectively and called CHIP1988, CHIP1995, CHIP2002, CHIP2007, and CHIP2013.

[16] Beijing and Shanghai were selected to represent China's large metropolitan cities; Liaoning, Jiangsu, Zhejiang, Fujian, and Guangdong to represent the eastern region; Shanxi, Anhui, Hebei, Henan, Hubei, and Hunan to represent the central region; and Chongqing, Sichuan, Yunnan, and Gansu to represent the western region. The provinces covered in the urban and rural surveys are almost identical, with the exception that Shanghai is only included in the urban survey and Hebei is only included in the rural survey.

[17] The data are derived from larger samples designed by China's State Statistics Bureau (SSB), but the questions about income are different from the SSB's surveys. Non-response is rare although excluded from the urban sample are those without a formal certificate of residence (*hukou*), an increasingly serious omission over time as the size of this population grows. Individuals are asked to keep a record of their incomes and

A considerable amount of time was spent in verifying the accuracy of the data and in 'cleaning' the data to eliminate measurement error. Essentially, this was affected by examining the sequence of income over the years for each individual and identifying unexpected or odd values. Sometimes these individuals were dropped from the sample. In other cases, it seemed reasonable to alter the recorded entry on income. This would occur when, for instance, zeros were missing in a single year.

The two main advantages of the survey for our study are: the quality of the data on income, earnings and expenditures (i.e. earnings is our outcome variable) and plethora of information on one's educational background, job networking, facets of one's social network, management responsibilities and leadership roles. Data on these domains is useful in examining various mechanisms underpinning the relationship our paper examines.

***The Chinese General Social Survey (CGSS).*** Our third data source is the Chinese General Social Survey (CGSS). The CGSS's main objective was to monitor systematically the changing relationship between social structure and quality of life in urban and rural China. The survey comprised urban households and 4,100 rural households in the 2003–6 Phase; the post-2006 design was slightly modified to recognize the changes in community development in rural and urban areas. The large sample size was required to reach larger sample sizes (1) for each of five strata that account for regional and geo-administrative variations of China, and (2) to allow an attrition and replacement rate of 15 percent between adjacent years of surveys.[18]

We use data from two CGSS years: 2003 and 2013. These years are conducted after the years of the CHIP survey and the provide additional labour market information to gauge the channels between party membership and earnings.

---

expenditures and they are asked to consult their records before providing information on incomes in previous years.

[18] The distribution of sampling units was designed as follows: (1) a total of 125 primary sampling units (PSU) were selected for the national sample; (2) four secondary sampling units (SSU) are selected in each selected PSU; (3) two third-level sampling units (TSU) are selected in each selected SSU; and (4) ten households are selected in each selected TSU. One eligible person age eighteen or over (eighteen to sixty-nine for the 2003 CGSS) was randomly selected from each sampled household to serve as the survey respondent. PSUs were county-level units. In official statistics, this refers to (a) counties (*xian*), (b) county-level cities (*xian ji shi*), and (c) city districts (*qu*) in cities whose administrative levels are prefecture or higher. Confined to the fifth population census, there are 2,801 PSUs from which 125 PSUs were selected by following these procedures: first, within each stratum, all PSUs are ranked according to the percentage of eligible respondents with a middle or higher educational level; and second, a given number of PSUs are selected by using a method of "proportionate to population size"; in this procedure, population refers to the civilian population ages eighteen to sixty-nine.

The CGSS survey offers our study three main advantages. First, the survey comprises very recent data on party membership, earnings and enables us to examine the same question as in the CHS and the CHIP surveys but with more recent data. Second, the CGSS survey collect data on parental party membership and so it provides us with an opportunity to apply an instrumental variable methodology using parental information as a potential instrument. Third, and perhaps the most important advantage of the survey relative to the previous two data sources, is the sub-sections that it covers from survey respondents on previous educational background, job attributes, overall happiness, job happiness, social network data and data on past job searches. Information on all of these variables is extremely helpful in examining potential mechanisms mediating the relationship between party membership and earnings.

**B.     Descriptive Statistics**

[Figure 1 about here] [Table 1-A, 1-B, 1-C, 1-D, 1-E about here]

Figure 1 reports data on the earnings distribution for the three surveys. Panel A reports the earnings distribution by party membership for CHIP 1988, Panel B reports the earnings distribution by party membership for CHS 1992, Panel C reports the earnings distribution by party membership for CHIP 2002, and the last two panels show the earnings distributions for CGSS 2003 and 2013.

In Table 1, we further detail the earnings summary statistics by party affiliation and the breakdown of various socio-economic factors by party affiliation. In these surveys, Communist party members make up anywhere from almost 10 percent to slightly more than 20 percent in the five survey samples. Communist party members tend to be slightly older than the general sample, with an average age in the mid-50s, compared to the sample average of high 40s. Communist party members are also disproportionately male compared to the entire sample. Communist party members are more likely to be married and to have attained some deree of higher education (i.e., college or above). Finally, the average monthly earnings of non-members are lower than the earnings of party members. Table 1 clearly shows a positive relationship between party membership and earnings.

## IV. Estimation Strategy

### A. Econometric Benchmark 1: Ordinary Least Squares

In general, the goal of this paper is to estimate an equation of the form:

(1) $\ln(Earn_i) = \beta_0 + \beta_1 C_i + \sum_{j=2}^{n} \beta_j X_{ji} + \theta_i + \delta_{ki} + \epsilon_i$

where $(Earn)_i$ represents monthly earnings, $C_i$, denotes if an individual is a member of the Communist party. $\sum_{j=2}^{n} \beta_j X_{ji}$ is a set of demographic variables: educational level[19], sex (whether or not a respondent is male), ethnicity (whether or not a respondent is of the Han majority), age (continuous definition), marital status (whether or not the respondent is married or not), religion (whether or not the respondent is religious or not), health (whether or not a respondent reports to be in poor health) and educational level attained. $\theta_i$ and $\delta_{ki}$ in (1) capture respectively the individual-specific and the district fixed effects for individual *i* living in district *k*.[20] Empirically, we cannot identify $\theta_i$ with a cross-sectional dataset (all three of our surveys) as there is only a single observation per person.

Our estimate of $\beta_1$ in (1) will capture the difference in earnings between a party member and non-party member, assuming that party membership $C_i$ is uncorrelated with other, unaccounted for factors that determine earnings. To gauge the magnitude and direction of the omitted variable bias, we can examine how $\beta_1$ changes as we add each additional term in (1).

### B. Econometric Benchmark 2: Propensity Score Matching

We augment the approach above with a propensity score matching method (Rosenbaum & Rubin 1983, 1984; Imbens 2004; Imbens & Wooldridge 2009; Imbens & Rubin 2014; Abadie *et al.* 2003; Angrist & Krueger 2000; Angrist & Pischke 2009).

---

[19] The dataset does not capture continuous years of education. Instead, the survey asks each respondent to report one of the following categories: "*No formal schooling*," "*Elementary*," "*Junior high school*," "*Senior high school*," "*Technical school*," "*Vocational school*," "*3 year college*," "*Formal college*," and "*Graduate school*."

[20] The districts in the city of Tianjin included in the dataset are Hepin, Nankai, Hexi, Hedong, Hongxiang, Hebei, Tanggu, Hanggu, and Dagang. The districts in the city of Shanghai included in the dataset are Huangpu, Nanshi, Luwan, Xuhui, Changning, Jingan, Putou, Zhabei, Hongkou, Yangpu, Minhang and Baoshan.

First, we estimate a propensity score for each observation, the likelihood of one being a Communist party member:

$$(2) \quad C_i = \alpha_0 + \sum_{j=1}^{n} \gamma_j \chi_{ji} + \epsilon_i$$

where $\sum_{j=1}^{n} \chi_{ji}$ is a vector of time-invariant variables: sex (whether or not a respondent is male), ethnicity (whether or not a respondent is of the Han majority), marital status, religion (whether or not the respondent is religious), and education attainment level.[21] The vector $\sum_{j=1}^{n} \chi_{ji}$ does not include variables that may have been affected by the treatment of interest (Rosenbaum, 1984; Frangakis & Rubin, 2002; Greenland, 2003).[22] Based on the propensity score, we then match individuals who are party members with their counterfactual units who are non-party members. Specifically, we use a 1:1 matching with replacement although for robustness we report other matching methods as well. In its simplest form, 1:1 nearest neighbor matching selects for each treated individual *i* the control individual with the smallest distance from individual *i*. In the final step we estimate the effect of Communist party membership on wages using:

$$(3) \quad Ln(Earn_i) = \beta_0 + \beta_1 \hat{C}_i + \sum_{j=2}^{n} \beta_j X_{ji} + \delta_{ki} + \epsilon_i$$

For the estimation of (2), we use only observations on the common support.[23] In the estimation procedure party members are matched with statistically similar (i.e., counterfactual) non-party members.

As a result, subclassifying or matching on the propensity score makes it possible to estimate treatment effects, controlling for covariates, because within subclasses that are homogeneous in the propensity score, the distributions of the covariates are the same for treated and control units (e.g., are "balanced"). In particular, for a specific value of the propensity score, the difference between the treated and control means for all units with that value of the propensity score is an unbiased estimate of the average treatment effect at that propensity score, assuming the conditional independence between treatment assignment and potential outcomes given the observed covariates ("strongly ignorable treatment assignment" assumption) (Rosenbaum and Rubin, 1983).

---

[21] Because educational levels are highly correlated with one's age, we do not include respondent's age though we estimate specifications with the age variable and the size of the key coefficient remains stable
[22] This is especially important when the covariates, treatment indicator, and outcomes are all collected at the same point in time, as is the case in our case.
[23] The common support ensures that persons with the same X values have a positive probability of being both participants and non-participants (Heckman, LaLonde, and Smith, 1999).

$\beta_1$ in specification (3) yields the *average treatment effect* of being a party member on one's monthly earnings. This assumption means that the treated and non-treated are similar in their observable characteristics and that selection into treatment is on observables only *i.e.* that there are no unobservable differences between the two groups that correlate with wages. Our matching approach and the various matching algorithms indeed capture all the relevant observable differences between those who are and are not in the Communist party. We address the identifying assumption below in the next two sections.

## V. Results: The Wage Premium of Party Membership

### A. Ordinary Least Squares

[Table 2 about here]

Table 2 reports the results from the OLS-based specification (1). The estimated coefficient is based on a regression with a full set of controls including regional fixed effects. The estimated earnings return to party membership ranged from 7.5 percent to almost 25 percent. Two facts merit attention. First, all of the estimated effect size on the communist party variable are highly statistically significant (at the 1-percent level). Second, the effect size based on these three survey sources seems to be increasing over the span of almost three decades. The estimated coefficient based on the CHIP 1988 survey is a 7.5 percent, implying that all else equal, being a party member increases one's earnings by 7.5 percent. The associated effect size but based on the CGSS 2013 sample is almost triple in size or 25 percent (statistically significant at the 1-percent). However, this empirical estimation is likely plagued from omitted variable bias (i.e., upward biased) and this is a concern we attempt to address below.

## B. Propensity Score Matching Results

To the extent party selection occurs (an issue we address below) on observable characteristics, we can also estimate the wage premium associated with being a Communist party member using the propensity score approach. Before we present the results based on this econometric approach, we provide analyses on the identifying assumptions.

***Common Support Assumption and Post-Matching Balancing.*** First, we test whether the *common support assumption* is fulfilled (Kahn and Tamer, 2010).[24] The substance of the common support assumption is that there must be both treated and untreated observations with each value of X.[25] The assumption essentially ensures that persons with the same X values have a positive probability of being both participants and non-participants (Heckman, LaLonde, and Smith, 1999). We examine graphically if the *common support* assumption holds in the five data samples. Figure 2 (and Appendix A Figure A.1) report the results.

[Figure 2 about here]

It is easy to discern, based on Figure 2, that in each class of the "propensity score" a certain number of "non-treated" individuals exist as well. Figure 3 displays the estimated density of the predicted probabilities that a communist party member is a non-member.[26] Based on Busso, DiNardo, and McCrary (2014) neither plot indicates too much probability mass near 0 or 1, and the two estimated densities have most of their respective masses in regions in which they overlap each other. Thus, there is no evidence that the overlap assumption is violated. Therefore, there is no visual evidence that the common support assumption is violated in these five data samples. To confirm the graphical test, we also perform a Kolmogorov-Smirnov (K-S) test (to test the equality of two distributions). The K-S test does not reject the null hypothesis of equality of distributions between groups after matching.

---

[24] The standard common support assumption is: $0 < \Pr(D=1|X) < 1$. The strict common support assumption is $0 < c < \Pr(D=1|X) < 1-c < 1$.
[25] When estimating the ATET, all that is required is untreated units for each value of X corresponding to at least one treated unit.
[26] Appendix A Figure A.1 reports the proportion of propensity scores by treatment status and it clearly shows an overlap between the distribution of propensity scores between treated and untreated units.

Next, we examine the balancing of covariates based on the propensity score matching exercise. Before we explore the effect of party membership on earnings, we examine balancing after the propensity matching has occurred.

[Figure 3 about here]

Related to the so-called *conditional independence* assumption, we assess the quality of matching after performing tests that check whether the propensity score adequately balances characteristics between the treatment and comparison group units. The objective of these tests is to verify that treatment is independent of unit characteristics after conditioning on observed characteristics (as estimated in the propensity score model).[27] It is important to note that only "after-matching" tests compare differences in time-invariant covariates (that are unaffected by treatment) for the resulting matched sample.

Figure 3 and Online Appendix A Table A.1 report the results of after-matching balancing. Figure 3 displays the overall balancing, based on the propensity score, for treated and untreated units. We examine graphically and with comparison of means to ensure that any differences in the covariate means between the two groups in the matched sample have been eliminated, which improves the likelihood of unbiased treatment effects. Appendix A Table A.1 reports the balancing post matching for each of the five datasets: CHIP 1988 is reported in Online Appendix A Table A.1-1, CHS 1993 is reported in Online Appendix A Table A.1-2, CHIP 2002 is reported in Online Appendix A Table A.1-3, CGSS 2003 is reported in Online Appendix A Table A.1-4, and CGSS 2013 is reported in Online Appendix A Table A.1-5. The tables also report a formal test for whether the matching is fulfilled by performing a formal for equality of means in the post-matching sample comparing the means between treated (i.e. communist party members) versus non-treated (non-party members). These formal tests reveal successful matching based on the chosen covariates. Furthermore, based on Imai and Ratkovic (2014), we conducted another formal balancing test for balancing of covariate means between treatment and control units (reported in Online Appendix A Table A.2).[28] The test fails to reject the null-hypothesis that the propensity score model is balanced based on the chosen covariate to predict party membership.

---

[27] Formally, this assumption entails $T \perp X \mid p(X)$, where X is the set of characteristics that are believed to satisfy the conditional independence assumption. In other words, after conditioning on $p(X)$, there should be no other variable that could be added to the conditioning set of the propensity score models that would improve the estimation, and after the application of matching, there should be no statistically significant differences between covariate means of the treatment and comparison units.

*Average Treatment Effects*. We estimate the average treatment effect (ATE) by propensity-score matching (PSM) based on a nearest neighbor 2:1 matching with replacement. PSM estimators impute the missing potential outcome for each subject by using an average of the outcomes of similar subjects that receive the other treatment level. Matching with replacement can often decrease bias because controls that look similar to many treated individuals can be used multiple times. This is particularly helpful in settings where there are few control individuals comparable to the treated individuals (e.g., Dehejia and Wahba, 1999). Additionally, when matching with replacement the order in which the treated individuals are matched does not matter. We return to the issue of the algorithm matching procedure and the robustness of the estimated effect size to the chosen algorithm choice in the next section.

[Table 3 about here]

Table 3 reports the results from the propensity score matching method based on specification (3).[29] The estimated effect sizes deserve several highlights. First, the estimated effect size range from approximately 18 percent based on the 1988 sample to 21 percent based on the CGSS sample. The CHS effect size is noticeably smaller although the CHS samples includes observations only in two specific urban areas (i.e., Tianjin and Shanghai), whereas the CHIP and CGSS samples include both urban and rural observations.[30] Second, the estimated effect size based on the matching procedure are generally lower compared to the effect size based on the OLS estimation. The decrease in the estimated effect size based on matching suggests that the matching procedure likely addresses additional positive selection that takes place in the party initiation process. We return, in Section 6, to the issue of how sensitive the effect size estimates are to the chosen matching algorithm used and the issue of potential additional bias based on the unobservable characteristics. Third, the reported effect size estimates show some evidence of an increase in the estimated wage premium associated with party membership.

---

[28] Imai and Ratkovic (2014) derive a test for whether the estimated propensity score balances the covariates. The score equations for parameters of the propensity-score model define an exactly identified generalized method of moments (GMM) estimator. Imai and Ratkovic (2014) use the conditions imposed by mean balance as over-identifying conditions. A standard GMM test for the validity of the over-identifying conditions is then a test for covariate balance.
[29] The PSM estimation was done using teffects in Stata 15.
[30] In analysis we do not report, we compare estimates from the same two regions in CGSS and the benchmark analysis shows extremely similar results.

### C.   Heterogeneous Treatment Analysis

Using the propensity score-based estimation, we examine how the treatment effect of party membership on earnings differ by important individual covariates. In particular, we focus on gender, education, ethnicity and whether a parent is a member of the Communist Party. To estimate the heterogeneous impact, we augment specification (3):

(4)  $Ln(Earn_i) = \beta_0 + \beta_1 \hat{C}_i + \sum_{j=2}^{n} \beta_j \hat{C}_i \times X_{ji} + \delta_{ki} + \epsilon_i$

$X_i$ captures covariates for which we test for treatment effect heterogeneity. Table 4 presents the combined effects (on the binary variable and the interaction).[31] We detect statistically significant differences for the Han ethnicity population. In terms of gender, we detect slightly larger effect sizes for women throughout the years our data allows to analyze.

[Table 4 about here]

### D.   Mechanisms

Next, we assess the relative importance of various channels in explaining the linkage between party membership and better wage outcomes. Party membership could change several aspects of daily life that could potentially contribute to the observed wage effects. In particular, it could improve the strength or the quality of one's social network, which could in turn confer various labour market benefits for one: reduce the job search time, provide information about available job opportunities, provide information on better-paying jobs. Party membership itself could also spur one's interest in getting better job qualifications (or certifications), which could carry intrinsic productivity benefits due to better knowledge acquisition or because they are explicitly valued and rewarded by employers. Third, party membership could confer access to better paying type of jobs. Fourth, party membership or better social capital could be an important determinant of one's overall well-being (Yip et al., 2007). Therefore, it seems likely that behavioural adjustments are (partly) responsible for the positive wage effects reported in the previous section. We investigate these mechanisms further with additional data on some of these potential channels.

---
[31] Conceptually the main difference between subgroup analysis and interaction terms is that stratified regressions allow all regression coefficients to vary across sub-groups; the difference between subgroup analysis and interaction term based regressions in practice will depend on number of control variables, and the assumption that control variables are orthogonal to the treatment.

We use the panel feature of the 2002 CHIP survey, which provides additional annual information on earnings and on potential mediating variables that could shed light of mechanisms underlying the relationship between party affiliation and one's wages. Based on the available data from the two survey sample, we explore for four main channels: strength of one's social network, human capital acquisition as one becomes a party member, improvement in one's social rank and overall life satisfaction. Specifically, the 2002 CHIP (the urban questionnaire) provides survey information on whether one holds a government job, how many friends one can rely on for finding a job, whether one has a professional title, months to find a new job, happiness level, and self-perceived social rank. Table 5 reports the results.

[Table 5 about here]

Column 1 reports the main results from Table 3. The remaining columns add the specific channel variables one at a time, and the last column (in Table 5) controls for all channel variables in the regression. In the final row, we only include the variables that are statistically significant in individual variable specifications and that do not entail a lot of missing observations. In the final regression in Table 5 (for CHIP 2002), we see strong evidence that the strongest impact on wage seems to be through an improvement the likelihood of one holding a government job, improvement in one's position within one's job hierarchy, and improvement in one's overall social rank. Although we see a positive effect size on the happiness variable, the variable is not statistically significant in the final specification. Of course, the evidence presented in Table 5 is only suggestive and relies on the assumption that the changes in the channel variables presented increase wage. Methodological problems and data limitations make it difficult to conduct a formal mediation analysis. Nevertheless, it does provide suggestive evidence that three important channels likely exert strong influence on one's wages: access to a subset of government jobs, the acquisition of additional job-related qualifications, and an overall improvement in one's social rank.

## VI. Robustness Checks

### A. Matching Algorithm Method

We examine robustness of the estimated effect sizes for the five data samples with respect to the matching algorithm method. In addition to the main matching results (which are based on the nearest neighbor 2:1 matching with replacement method), we further re-estimate the effect sizes using the nearest neighbour matching (NN), the caliper method, the kernel method, and the IPW matching method. The most straightforward matching estimator is nearest neighbour (NN) matching. The individual from the comparison group is chosen as a matching partner for a treated individual that is closest in terms of propensity score. Several variants of NN matching are possible, e.g. NN matching "with replacement" and "without replacement". In the former case, an untreated individual can be used more than once as a match, whereas in the latter case it is considered only once. Matching with replacement involves a trade-off between bias and variance. NN matching faces the risk of bad matches, if the closest neighbour is far away. This can be avoided by imposing a tolerance level on the maximum propensity score distance (i.e., caliper matching). Imposing a caliper works in the same direction as allowing for replacement. Bad matches are avoided and hence the matching quality rises. The idea of stratification matching is to partition the common support of the propensity score into a set of intervals (strata) and to calculate the impact within each interval by taking the mean difference in outcomes between treated and control observations (Rosenbaum and Rubin 1983). We use five subclasses, purported to remove 90 percent of the bias due to measured confounders, have been used by the majority of propensity score studies (Thoemmes & Kim 2010) based on Cochran (1968) and Rosenbaum and Rubin (1984).

Kernel matching (KM) and local linear matching (LLM) are non-parametric matching estimators that use weighted averages of all individuals in the control group to construct the counterfactual outcome. Thus, one major advantage of these approaches is the lower variance which is achieved because more information is used. Imbens (2004) notes that propensity scores can also be used as weights to obtain a balanced sample of treated and untreated individuals (IPW method). If the propensity score is known, the estimator can directly be implemented as the difference between a weighted average of the outcomes for the treated and untreated individuals.

In Online Appendix B Tables B.2, we present estimates from the algorithm matching techniques outlined above. The results in the table show that the effect size estimates are fairly robust to the choice of algorithm matching technique – estimates range from 8.25 to 9.24 percent higher wage premium for communist party membership. Only the radius caliper (0.2) matching for CGSS 2013 yields a slightly lower premium.[32] [33]

### B. Two-Stage Least Squares Method

We also augment our matching procedure results with a two-stage least squares estimation (2SLS), in which we instrument one's party affiliation with one's parental party affiliation. A valid instrument, such as parental party membership, for one's party affiliation must shift one's party affiliation but affect one's monthly earnings only through one's own party membership. Although this estimation approach is a promising alternative with several advantages to the matching procedure, only a subset of the datasets we use have data on parental party affiliation and even for these datasets, data on parental affiliation is missing for a very large number of the observations. Based on the two datasets (CHIP 2002 and CGSS 2013), we re-estimate the wage premium associated with party premium with the limited data we have on parental party affiliation in these two data samples.

Appendix A.3 reports the results for the estimated wage premium. Appendix A.3, Column (1) reports the estimated effect size for the CHIP 2002 dataset and the estimated effect size is a 17 percent increase (imprecisely estimated) in monthly earnings associated with Communist party membership. Appendix A.3, Column (2) reports the estimated effect size for the CGSS 2002 dataset and the estimated effect size is a 27 percent increase in monthly earnings (imprecisely estimated) associated with Communist party membership. Although both estimates of the wage premium coefficient are not statistically significant, they are comparable to the results based on the matching procedure. The CGSS 2013 based on the 2SLS is slightly higher than the effect size based on the matching, which could be due to the fact that the instrument cleanses some measurement error in the outcome variable and/or because the 2SLS estimate is based on a very different subset of the sample population, the-so called group of *compliers* (Angrist and Krueger, 1999), which could be the subset of individuals with higher marginal return to party membership to begin with.

---

[32] We also explore the stability of the results for the NN matching technique by the number of propensity influencing variables (not reported and available upon request). To satisfy the assumption of ignorable treatment assignment, it is important to include in the matching procedure all variables known to be related to both treatment assignment and the outcome (Rubin and Thomas, 1996; Heckman *et al.*, 1998; Glazerman *et al.*, 2003; Hill *et al.*, 2004). We explore adding five to fifteen additional variables and examine how the treatment effects change. The results are stabile to the inclusion of additional matching variables.

[33] We explore in Online Appendix Tables B.1-1 through B.1-5 how treatment varies by the probability of selection into treatment based on Xie, Brand, and Jann (2012). In Online Appendix Table B.1-1 to B-1.5, we examine variation in the effect size by the treatment probability. We detect very small differences by propensity score strata. The results regarding the relationship between party membership and monthly earnings are stable and most differences are statistically insignificant.

### C. Quantile Regressions

We also estimate (reported in Online Appendix B Tables B.3) the effect of party membership by estimating an equation expressing each quantile of the conditional distribution. In this type of estimation, we allow for effects of the independent variables to differ over the quantiles. We estimate the propensity score method for each quantile – 0.25, 0.50 and 0.75. Using these specifications, we find that for individuals who earn less (i.e., 0.25 quantile), the effect of party membership on earnings is particularly pronounced. We cannot detect strong effects for individuals when we restrict our attention to the 0.50 and 0.75 quantiles. We also graph (in Online Appendix A Figure A.2) the estimates coefficients for the effect of being a communist party member on wages for each quantile regression based on Koenker and Basset (1978) with quantile increments of 0.05.

### D. Rosenbaum Bounds

The estimation of treatment effects relies on the matching estimators is based on the *conditional independence* assumption (CIA), i.e. selection on observable characteristics. If there are unobserved variables which affect assignment into treatment and the outcome variable simultaneously, a *hidden bias* might arise. In this section, we explore how sensitive the treatment effect is if inference about treatment effects is altered by unobserved factors. We examine how strongly an unmeasured variable must influence the selection process in order to undermine the implications of matching analysis presented. Rosenbaum (2002) has developed a method of sensitivity analysis to assess if one's estimated based on matching is robust to the possible presence of an unobserved confounder. This sensitivity analysis for matched data provides a specific statement about the magnitude of hidden bias that would need to be present to explain the associations actually observed (Rosenbaum, 2002).

We estimate the *Rosenbaum bounds* based on the main estimation matching technique. The results are reported in Online Appendix B Tables B.4-1 through B.4-5. $\Gamma$ is a measure of the degree of departure from a study that is free of bias. Overall the lowest critical value for $\Gamma$ (gamma) ranges from 1 to 10 and varies between the Hodges-Lehmann point estimate and the 95 percent confidence interval.[34] Gamma captures the log odds of differential assignment due to unobserved factors. In other words, gamma allows us to examine if we introduce differential likelihood for assignment into treatment group, how the treatment effect will change. We find that the lowest critical value that (barely) includes zero ranges from 2.00 to 5.00 (Hodges-Lehmann point estimate). Such a high H-L critical value constitutes strong evidence that our estimated positive effects of the effect of communist party membership on wages are robust to even small amount of bias based on selection on unobservables.

## VII. Conclusions

One million Chinese citizens join the Communist party of China every year, and over 80 percent of graduating college students apply. Membership in the party is perceived as an investment in political capital that can help one get a better job and higher salaries. In this paper, we estimate the wage premium of membership in the Chinese Communist party using data from over three decades.

We report three main findings. First, using a propensity score matching technique, we find that Communist party members, on average, earn approximately a little over 20 percent more in monthly earnings compared to non-members. This estimated effect is larger than previous estimates summarized in Li et al. (2007) – our estimates are based on sample from both rural and urban areas in China, whereas previous studies rely on data predominantly from urban areas. This finding adds to previous research using data from developing countries that shows substantial monetary benefits associated with political status and social connections (Siddique, 2010; Madheswaram and Attewell, 2007; Das and Dutta, 2007). To bolster the credibility of our estimates, we examine the robustness of the results with respect to various factors: the matching algorithm method, the estimation technique and potential selection bias due to unobservable characteristics. Second, because we rely on data over three decades, we find suggestive evidence that this wage premium has grown modestly over the last three decades. In both the ordinary least squares and propensity score matching results, we detect evidence that the wage premium has increased. Finally, we explore for the relative importance of various channels in explaining the linkage between party membership and better wage outcomes. Based on the available data from the two data samples, we explore for four main channels: strength of one's social network, human capital acquisition as one becomes a party member, improvement in one's social rank and overall life satisfaction. We provide suggestive evidence that at least three important channels likely exert very strong positive influence on one's wages when one joins the Communist Party: better access to government jobs, improvement in one's position within one's job hierarchy, and an overall improvement in one's social rank.

---

[34] See Hollander and Wolfe (2013) for more details on the Hodges-Lehmman point estimate for the sign rank test.

All of the findings provide robust evidence that political connections can play an important economic role in the world's most populous economy. The results in this paper also shed new light one reason that helps explain why the communist party membership has more than doubled since the early 1980s and is likely to continue to do so in future years.

# Figures

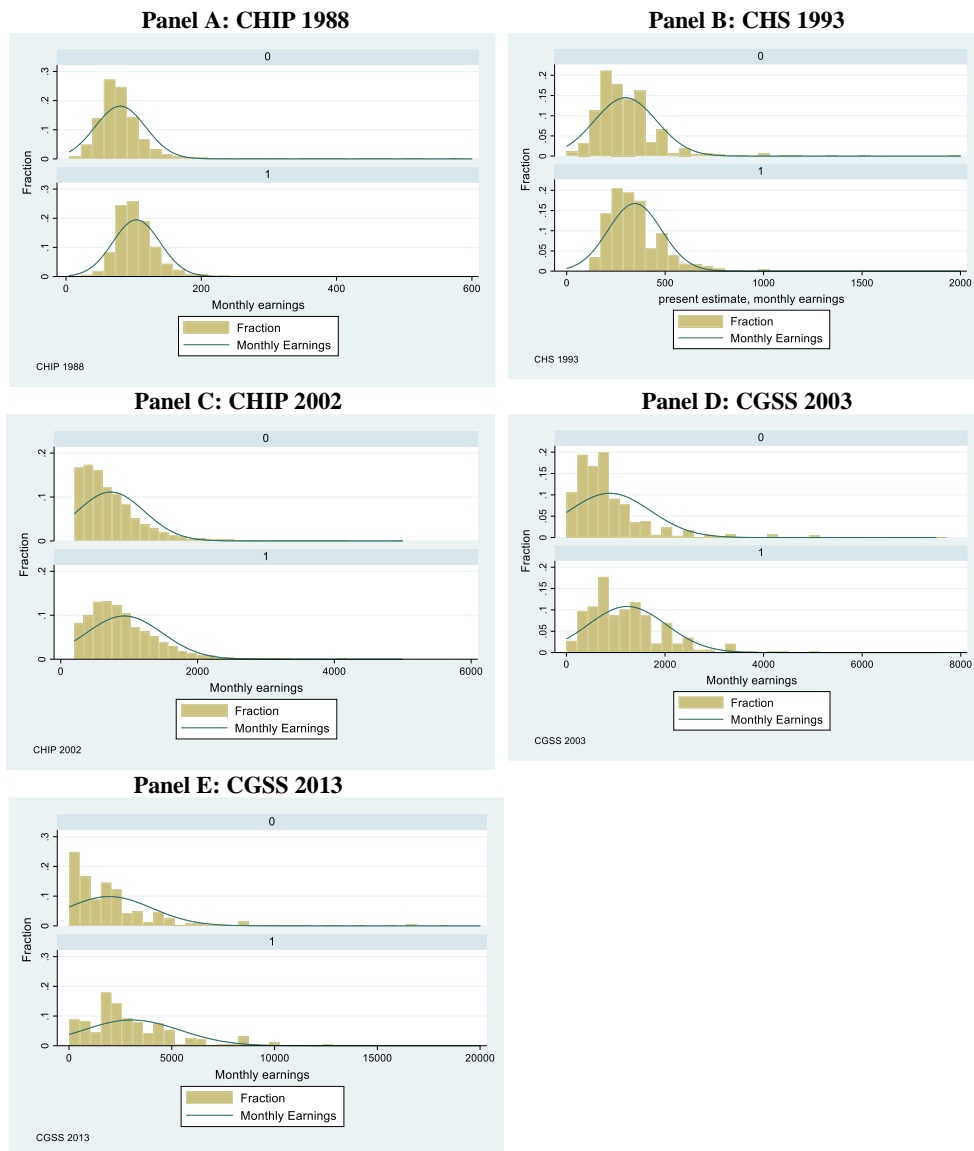

**FIGURE 1** Distribution of Logged Monthly Earnings (in RMB), By Survey Source
Note: Distribution for each value of the party membership variable (1=Communist party member; 0=non-member)

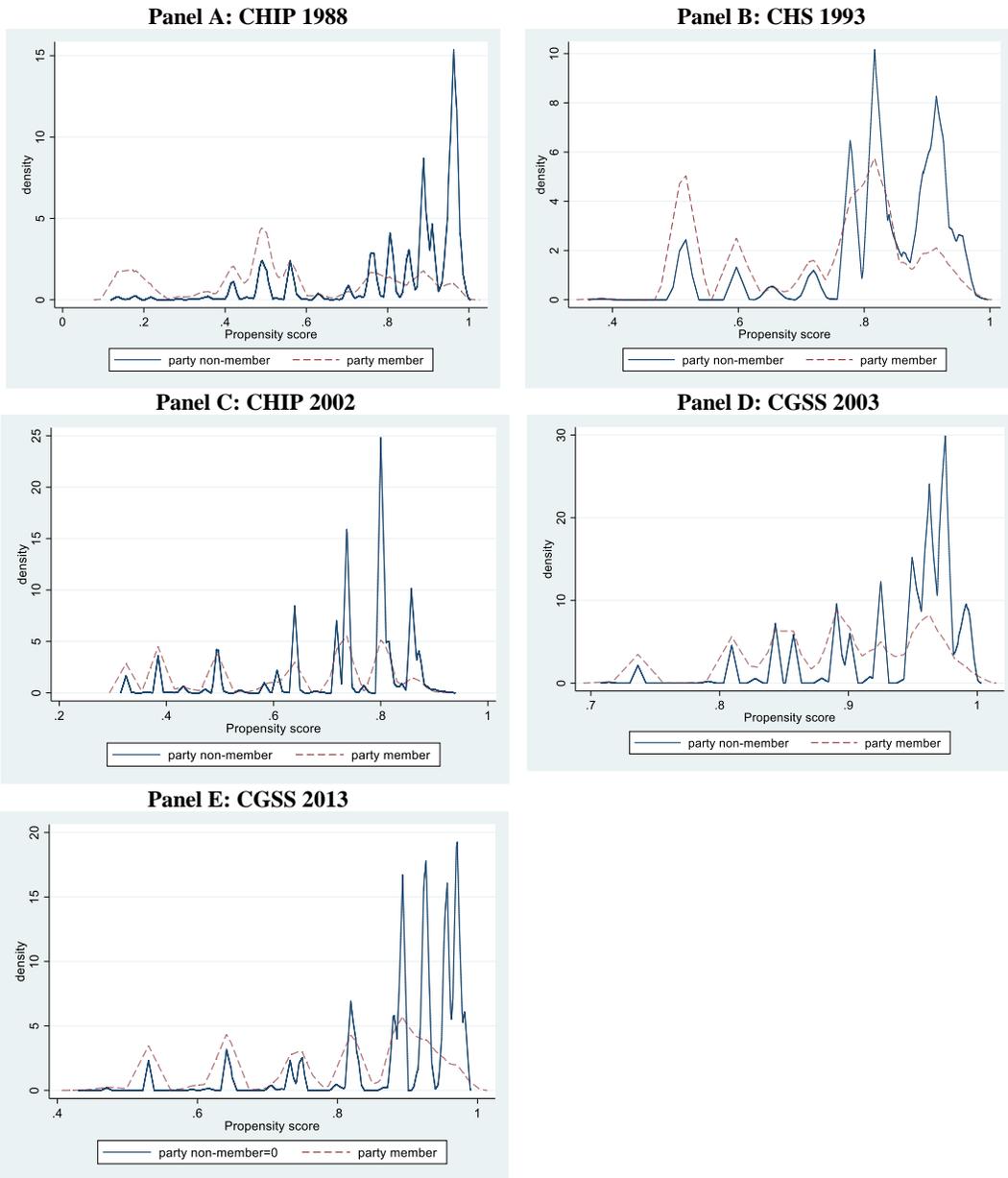

**FIGURE 2** Density of the Predicted Probability

Note: Shows density distributions of participants and non-participants, and the region of common support; X-axis: *high* probability of participating given X

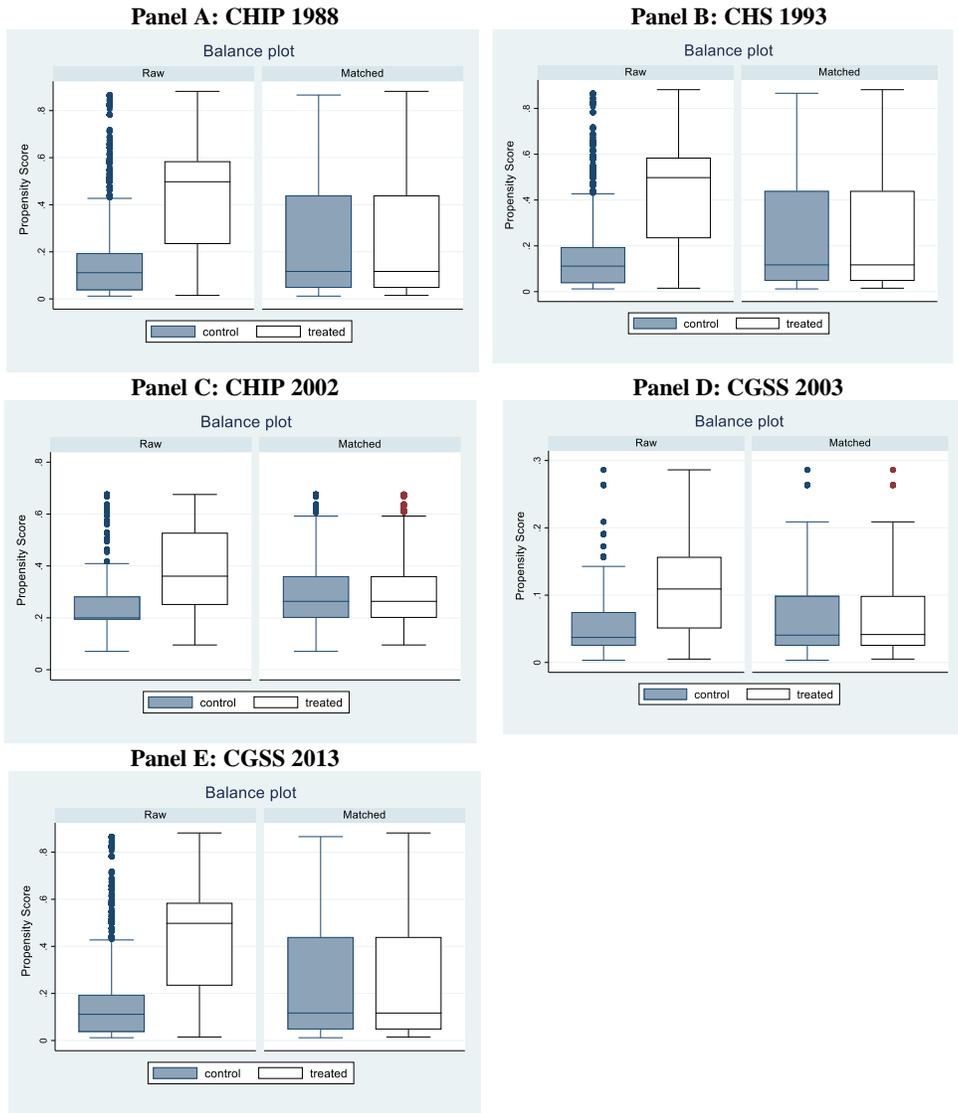

**FIGURE 3** Balancing Post Propensity Score Matching
Note: Distribution for each value of the party membership variable (1=Communist party member; 0=non-member)

# Tables

| | | Sample | Non-Communist Party Members | Communist Party Members |
|---|---|---|---|---|
| | | (1) | (2) | (3) |
| Monthly Earnings (in Rmbs) | | 87.41 | 82.01 | 105.31 |
| | | (57.73) | (58.46) | (51.91) |
| Member in the Communist party (percent) | | 12.50% | 0% | 100% |
| Han (percent) | | 94.00% | 93.87% | 94.89% |
| Male (percent) | | 49.69% | 46.85% | 79.43% |
| Age | | 39.80 | 37.04 | 47.36 |
| | | (51.71) | (37.92) | (42.19) |
| Education Level | Primary | 5.44% | 6.27% | 1.79% |
| | Middle School | 13.26% | 14.54% | 7.69% |
| | High School | 35.45% | 37.51% | 26.41% |
| | Technical School | 21.28% | 21.81% | 18.91% |
| | Vocational School | 2.86% | 3.17% | 1.54% |
| | College[a] | 7.40% | 6.27% | 12.31% |
| Occupation | Private Sector | 1.00% | 1.12% | 0.99% |
| | Professional | 4.89% | 4.32% | 15.21% |
| | Managerial | 3.18% | 1.28% | 20.41% |
| | Office | 7.03% | 5.71% | 25.46% |
| | Low-skill | 18.33% | 22.65% | 11.80% |
| | Agricultural | 44.14% | 55.06% | 17.94% |
| | Temporary | 1.24% | 1.58% | 0.00% |
| Observations | | 60,897 | 45,338 | 6,476 |

**TABLE 1-A** Descriptive Statistics (CHIP 1988)

Notes: Standard errors in parentheses. [a]College combines individuals who reported having attained college-level education and graduate school.

**TABLE 1-B** Descriptive Statistics (CHS 1993)

|  |  | Sample | Non-Communist Party Members | Communist Party Members |
|---|---|---|---|---|
|  |  | (1) | (2) | (3) |
| Monthly Earnings (in Rmbs) |  | 307.77 (155.093) | 298.59 (157.93) | 346.21 (136.24) |
| Member in the Communist party (percent) |  | 18.61% | 0% | 100% |
| Han (percent) |  | 98.47% | 98.53% | 98.21% |
| Male (percent) |  | 60.31% | 56.15% | 78.46% |
| Age |  | 47.39 (13.76) | 46.50 (14.03) | 51.27 (11.79) |
| Religious (percent) |  | 4.3% | 4.76% | 2.31% |
| Married (percent) |  | 86.07% | 84.76% | 91.79% |
| In Poor Health (percent) |  | 10.16% | 9.96% | 11.03% |
| Education Level | Elementary | 5.44% | 6.27% | 1.79% |
|  | No schooling | 13.26% | 14.54% | 7.69% |
|  | Elementary | 35.45% | 37.51% | 26.41% |
|  | Middle School | 21.28% | 21.81% | 18.91% |
|  | High School | 2.86% | 3.17% | 1.54% |
|  | Technical School | 7.40% | 6.27% | 12.31% |
|  | Vocational School | 8.78% | 6.39% | 19.23% |
|  | Three-year College | 5.34% | 3.93% | 11.54% |
|  | Formal College | 0.14% | 0.12% | 0.26% |
|  | Graduate School | 5.44% | 6.27% | 1.79% |
| Observations |  | 2,096 | 1,621 | 390 |

Notes: Standard errors in parentheses

**TABLE 1-C** Descriptive Statistics (CHIP 2002)

|  |  | Sample | Non-Communist Party Members | Communist Party Members |
|---|---|---|---|---|
|  |  | (1) | (2) | (3) |
| Monthly Earnings (in Rmbs) |  | 593.98 | 525.09 | 835.22 |
|  |  | (562.07) | (527.99) | (610.77) |
| Member in the Communist party (percent) |  | 20.40% | 0% | 100% |
| Han (percent) |  | 94.00% | 93.87% | 94.89% |
| Male (percent) |  | 49.69% | 46.85% | 79.43% |
| Age |  | 39.80 | 37.04 | 47.36 |
|  |  | (51.71) | (37.92) | (42.19) |
| Married (percent) |  | 76.61% | 72.46% | 94.50% |
| Education Level | Primary | 18.34% | 19.47% | 10.17% |
|  | Middle School | 54.71% | 56.12% | 44.95% |
|  | High School | 14.01% | 12.06% | 27.92% |
|  | Technical School | 5.04% | 4.49% | 8.91% |
|  | Vocational School | 1.46% | 1.13% | 3.86% |
|  | College[a] | 0.28% | 0.22% | 0.71% |
| Occupation | Private Sector | 3.83% | 4.26% | 2.25% |
|  | Professional | 8.45% | 7.22% | 13.21% |
|  | Managerial | 4.47% | 1.63% | 15.68% |
|  | Office | 9.91% | 6.76% | 22.33% |
|  | Low-skill | 8.38% | 9.06% | 5.93% |
|  | Agricultural | 9.45% | 11.17% | 3.09% |
|  | Temporary | 7.14% | 8.09% | 3.75% |
| Observations |  | 60,897 | 45,338 | 6,476 |

Notes: Standard errors in parentheses. [a]College combines individuals who reported having attained college-level education and graduate school.

**TABLE 1-D** Descriptive Statistics (CGSS 2003)

| | | Sample | Non-Communist Party Members | Communist Party Members |
|---|---|---|---|---|
| | | (1) | (2) | (3) |
| Monthly Earnings (in Rmbs) | | 941.60 | 917.31 | 1,294.82 |
| | | (298.40) | (1,080.23) | (1,093.38) |
| Member in the Communist party (percent) | | 6.44% | 0% | 100% |
| Han (percent) | | 94.43% | 94.46% | 94.12% |
| Male (percent) | | 53.57% | 54.07% | 46.37% |
| Age | | 44.62 | 44.13 | 51.74 |
| | | (12.69) | (12.62) | (11.47) |
| Married (percent) | | 84.97% | 84.27% | 95.16% |
| Education Level | Primary | 12.78% | 13.21% | 6.57% |
| | Middle School | 30.55% | 31.75% | 13.15% |
| | High School | 18.70% | 18.87% | 16.26% |
| | Technical School | 10.00% | 9.54% | 16.61% |
| | Vocational School | 3.14% | 3.31% | 0.69% |
| | College | 6.80% | 5.93% | 19.38% |
| | Graduate School | 0.49% | 0.48% | 0.69% |
| Observations | | 4,491 | 4,202 | 289 |

Notes: Standard errors in parentheses

**TABLE 1-E** Descriptive Statistics (CGSS 2013)

| | | Sample | Non-Communist Party Members | Communist Party Members |
|---|---|---|---|---|
| | | (1) | (2) | (3) |
| Monthly Earnings (in Rmbs) | | 2,240.94 | 2,082.62 | 3,437.46 |
| | | (3,165.13) | (2,818.19) | (4,909.80) |
| Member in the Communist party (percent) | | 11.69% | 0% | 100% |
| Han (percent) | | 91.35% | 91.25% | 92.08% |
| Male (percent) | | 54.58% | 51.92% | 74.72% |
| Age | | 49.40 | 48.88 | 53.26 |
| | | (15.72) | (15.57) | (16.29) |
| Married (percent) | | 78.82% | 78.23% | 83.30% |
| Education Level | Primary | 21.63% | 23.27% | 9.25% |
| | Middle School | 30.00% | 31.36% | 19.72% |
| | High School | 11.52% | 11.31% | 13.11% |
| | Technical School | 2.28% | 2.37% | 16.04% |
| | Vocational School | 5.27% | 4.81% | 8.77% |
| | College | 7.22% | 5.31% | 21.70% |
| | Graduate School | 0.68% | 0.46% | 2.36% |
| Observations | | 9,071 | 8,011 | 1,060 |

Notes: Standard errors in parentheses

**TABLE 2** Earnings Equation (OLS)

| Dependent Variable: | ln (Monthly Earnings), (in RMBs) | | | | |
|---|---|---|---|---|---|
| Survey Source /Year | CHIP 1988[a] | CHS 1993[b] | CHIP 2002[c] | CGSS 2003[d] | CGSS 2013[e] |
| | (1) | (2) | (3) | (4) | (5) |
| Communist Party Membership | 0.075*** (0.006) | 0.134*** (0.035) | 0.163*** (0.023) | 0.171*** (0.047) | 0.253*** (0.032) |
| District Fixed Effects | YES | YES | YES | YES | YES |
| R-squared | 0.369 | 0.359 | 0.3244 | 0.3118 | 0.4272 |
| Observations | 19,323 | 1,994 | 11,817 | 4,491 | 9,071 |

Notes: Standard errors in parentheses. [a] in this specification, the control variables are educational level, ethnicity, gender, age, age-sq, urbanicity, religious status, marital status, health status; [b] in this specification, the control variables are educational level, ethnicity, gender, age, age-sq, urbanicity; [c] in this specification, the control variables are educational level, ethnicity, gender, age, age-sq, urbanicity, marital status, health status. [d] in this specification, the control variables are educational level, ethnicity, gender, age, age-sq, urbanicity, marital status, health status; [e] in this specification, the control variables are educational level, ethnicity, gender, age, age-sq, urbanicity, marital status, health status. ***, ** and * indicate significance at 1, 5 and 10%, respectively.

**TABLE 3** Earnings Equation (Propensity Score Matching Estimation)

| Dependent Variable: | ln (Monthly Earnings), (in RMBs) | | | | |
|---|---|---|---|---|---|
| Survey Source /Year | CHIP 1988[a] | CHS 1993[b] | CHIP 2002[c] | CGSS 2003[d] | CGSS 2013[e] |
| | (1) | (2) | (3) | (4) | (5) |
| Communist Party Membership | 0.177*** | 0.094*** | 0.188** | 0.233*** | 0.212*** |
| | (0.010) | (0.024) | (0.022) | (0.032) | (0.052) |
| District Fixed Effects | YES | YES | YES | YES | YES |
| R-squared | 0.369 | 0.359 | 0.401 | 0.3118 | 0.4272 |
| Observations | 19,323 | 1,994 | 11,817 | 4,491 | 9,071 |

Notes: Standard errors in parentheses. [a] in this specification, the control variables are educational level, ethnicity, gender, age, age-sq, urbanicity, religious status, marital status, health status; [b] Variables used in estimating propensity score: sex dummy (1 if respondent is male and 0 if female), ethnicity dummy (1 if respondent is of the Han majority ethnicity and 0 if of a minority ethnicity), married dummy (1 if the respondent is married and 0 if single), dummies for education (1 if the respondent achieved the specified level of education and 0 if not) and religion (1 if the respondent is religious and 0 if town). Values can be interpreted as the percent change in monthly earnings. Communist party membership in a dummy equal to 1 if the subject is a member of the Communist party and 0 if otherwise; [c] in this specification, the control variables are educational level, ethnicity, gender, age, age-sq, urbanicity, marital status, health status. [d] in this specification, the control variables are educational level, ethnicity, gender, age, age-sq, urbanicity, marital status, health status; [e] in this specification, the control variables are educational level, ethnicity, gender, age, age-sq, urbanicity, marital status, health status.

***, ** and * indicate significance at 1, 5 and 10%, respectively.

**TABLE 4** Heterogeneous Treatment Analysis

| Socio-economic Group | | ln (Monthly Earnings, in RMBs) | | | | |
|---|---|---|---|---|---|---|
| | | CHIP 1988 | CHS 1993 | CHIP 2003 | CGSS 2003 | CGSS 2013 |
| | | (1) | (2) | (3) | (4) | (5) |
| Gender | Male | 0.161*** (0.010) | 0.124*** (0.029) | 0.197*** (0.027) | 0.211*** (0.035) | 0.106** (0.05) |
| Education | College Degree | 0.142*** (0.020) | -0.075 (0.078) | 0.082*** (0.054) | -0.114 (0.102) | -0.025 (0.065) |
| Ethnicity | Han Ethnicity | 0.181*** (0.011) | 0.093*** (0.024) | 0.184*** (0.022) | 0.226*** (0.033) | 0.200*** (0.055) |
| Parent Communist | Parental Communist Party Member Status | NA | -0.020 (0.065) | 0.134*** (0.038) | NA | NA |
| Estimation Strategy | | PSM | PSM | PSM | PSM | PSM |
| Observations | | 19,323 | 1,994 | 24,704 | 4,491 | 9,071 |

Notes: ***, ** and * indicate significance at 1, 5 and 10%, respectively.

**TABLE 5** Mechanisms of Party Influence on Earnings (CHIP 2002[a])

| Variables | Dependent Variable: ln (Monthly Earnings), (in RMBs) | | | | | | | | |
|---|---|---|---|---|---|---|---|---|---|
| | (1) | (2) | (3) | (4) | (5) | (6) | (7) | (8) | (9)[h] |
| CPM[a] | 0.188** | 0.152 | 0.167 | 0.185 | 0.210* | 0.210* | 0.237*** | 0.244** | 0.079 |
| | (0.022) | (0.120) | (0.119) | (0.119) | 0.120 | 0.120 | (0.096) | (0.096)) | (0.120) |
| Holds a government job (1=yes) | | 0.067*** | | | | | | | 0.599* |
| | | (0.027) | | | | | | | (0.155) |
| Higher professional title (1=yes)[c] | | | 0.471*** | | | | | | 0.324* |
| | | | (0.167) | | | | | | (0.170) |
| Friends who can help one find a job? (#)[d] | | | | 0.007 | | | | | |
| | | | | (0.023) | | | | | |
| Holds a management position | | | | | 1.225 | | | | |
| | | | | | (0.852) | | | | |
| Months to find a job? (#)[e] | | | | | | 0.004 | | | |
| | | | | | | (0.027) | | | |
| Happiness level[f] | | | | | | | 0.215*** | | 0.085 |
| | | | | | | | (0.049) | | (0.061) |
| Self-perceived social rank[g] | | | | | | | | 0.490*** | 0.336*** |
| | | | | | | | | (0.065) | (0.084) |
| Observations | 5,825 | 4,115 | 4,140 | 4,140 | 4,114 | 222 | 5,768 | 5,811 | 1,961 |

Notes: (a) For this analysis we only use the urban sub-sample of CHIP 2002 because the survey questions on these potential mechanisms are only available in the urban survey questionnaire. (b) CPM=Communist Party Member. (c) Professional title and administrative rank of professionals and cadres of government agents, institutions and enterprises. Coded as 1 if individual reported having a senior title, being a bureau chief level and above, or division chief level and above, or section chief level and above. (d) The survey question was "If you want to change your job, how many friends and relatives can you ask to help you?" Robust standard errors in parentheses. (e) variable is not included in the final regression in column (6) because of very high number of missing observations that results in a very small sample size for that specification. (f) Happiness level is five levels and, in our regressions, the higher value indicates happier individual. The levels are: very happy, happy, so-so, not very happy, not happy at all. (g) self-perceived social rank based on living standards. Higher values indicate increase in social rank. The actual categories are: bottom quartile, second lowest quartile, second best quartile, and top quartile. (h) in this specification, we only include the variables that have do not have a considerable number of missing observations. ***, ** and * indicate significance at 1, 5 and 10%, respectively.

# Online Appendix A

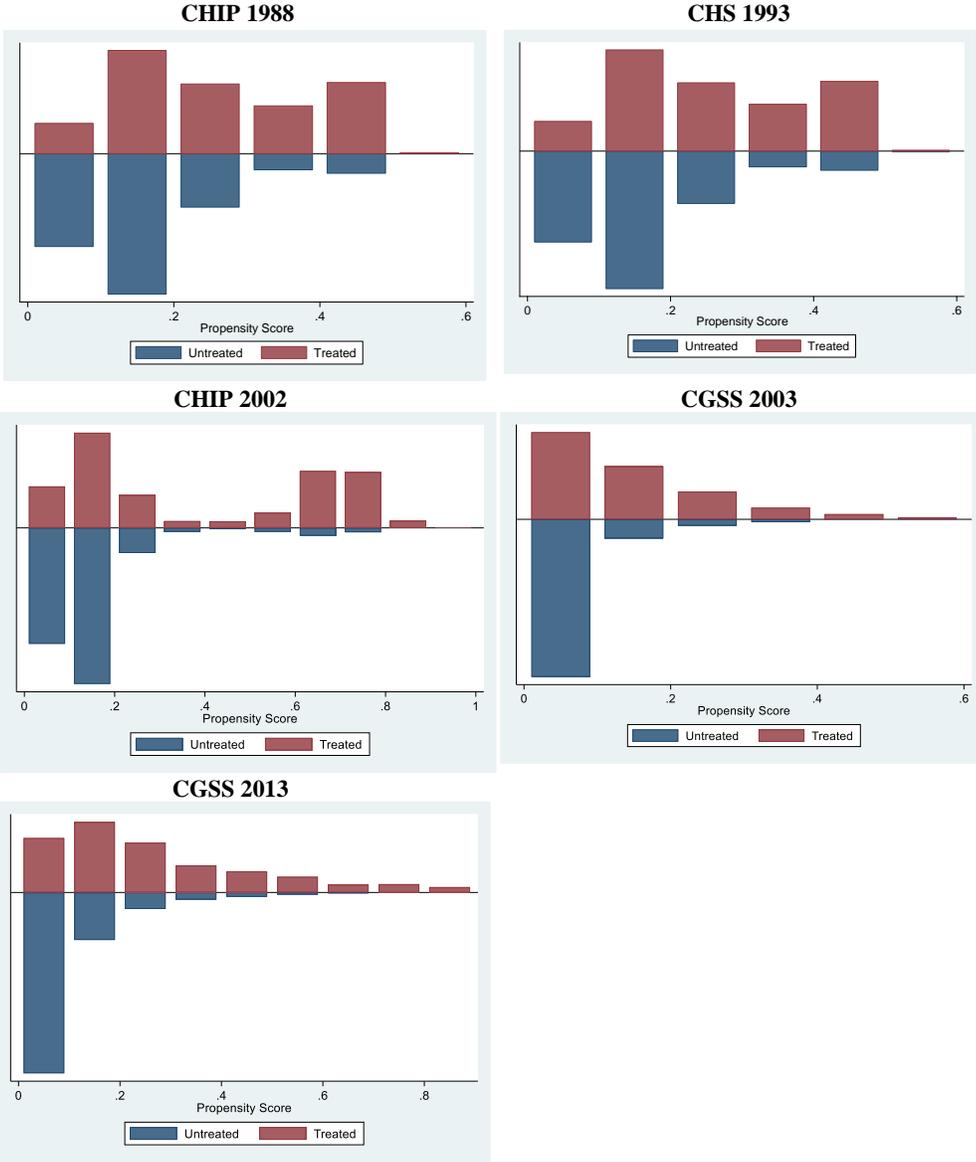

**FIGURE A.1** Visual Check for Common Support Assumption

Note: Graphical check that the "common support" assumption is fulfilled. The assumption is fulfilled when there is sufficient overlap between the distributions of propensity scores across treatment and control groups. the y axis in psgraph is proportional by group – the treated and untreated are not necessarily on the same scale. Performed in Stata 15 with psgraph.

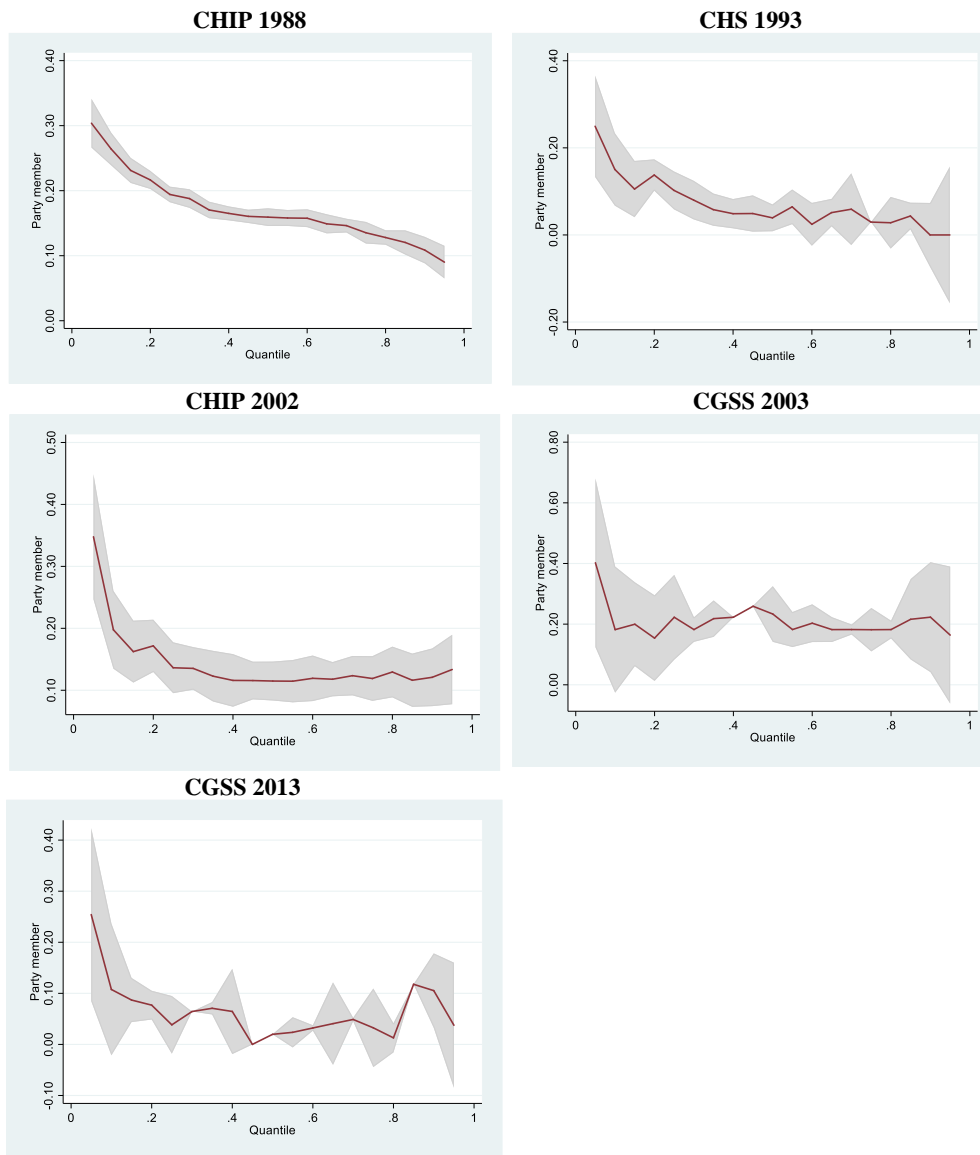

**FIGURE A.2** Quantile Regression Confidence Intervals

Note: The figure displays the coefficients of a quantile regression (Koenker and Basset, 1978) and also reports the OLS confidence interval.

**TABLE A.1-1** Balancing Post Matching (CHIP 1988)

|  | Pre-matched Means (Variance) | | Post-matched Means (Variance) | | $t$-value for matched sample |
|---|---|---|---|---|---|
|  | Treatment | Control | Treatment | Control |  |
| Han | 0.954 (0.044) | 0.966 (0.033) | 0.954 (0.044) | 0.966 (0.033) | -0.88 |
| Male | 0.775 0.174) | 0.469 (0.249) | 0.775 (0.174) | 0.469 (0.249) | -0.05 |
| Primary School | 0.087 (0.079) | 0.126 (0.110) | 0.126 (0.110) | 0.036 (0.035) | -0.07 |
| Middle School | 0.298 (0.209) | 0.420 (0.244) | 0.123 (0.108) | 0.042 (0.040) | 0.02 |
| High School | 0.196 (0.158) | 0.256 (0.190) | 0.154 (0.131) | 0.088 (0.081) | -0.05 |
| College | 0.126 (0.110) | 0.036 (0.035) | 0.196 (0.158) | 0.256 (0.190) | -0.51 |
| Vocational School | 0.123 (0.108) | 0.042 (0.040) | 0.298 (0.209) | 0.420 (0.244) | 0.55 |
| Technical School | 0.154 (0.131) | 0.088 (0.081) | 0.087 (0.079) | 0.126 (0.110) | -0.20 |
| Private Sector | 0.010 (0.010) | 0.007 (0.07) | 0.010 (0.010) | 0.007 (0.007) | 1.12 |
| Professional | 0.215 (0.169) | 0.125 (0.110) | 0.215 (0.169) | 0.125 (0.110) | -0.21 |
| Director | 0.249 (0.187) | 0.025 (0.024) | 0.249 (0.187) | 0.025 (0.024) | 0.00 |
| Office | 0.357 (0.230) | 0.169 (0.141) | 0.357 (0.230) | 0.169 (0.141) | -0.00 |
| Manual Labor | 0.154 (0.130) | 0.630 (0.223) | 0.154 (0.130) | 0.630 (0.223) | 0.00 |
| Agricultural | 0.010 (0.010) | 0.020 (0.019) | 0.010 (0.010) | 0.020 (0.019) | 0.00 |
| Temporary | 0.003 (0.003) | 0.018 (0.018) | 0.003 (0.003) | 0.018 (0.018) | -0.00 |
| Urban Location | 0.930 (0.065) | 0.889 (0.099) | 0.930 (0.065) | 0.889 (0.099) | 0.17 |
| Observations | 4,494 | 14,829 | 4,494 | 14,829 |  |

Notes: Variances in parentheses. (a) t-tests for equality of means in the two samples (before and after matching if option both is specified). T-tests are based on a regression of the variable on a treatment indicator. Before matching or on raw samples this is an unweighted regression on the whole sample, after matching the regression is weighted using the matching weight variable _weight or user-given weight variable in mweight and based on the on-support sample. T-tests are based on Rosenbaum and Rubin (1985).

**TABLE A.1-2** Balancing Post Matching (CHS 1993)

|  | Pre-matched Means (Variance) | | Post-matched Means (Variance) | | t-value for matched sample |
|---|---|---|---|---|---|
|  | Treatment | Control | Treatment | Control |  |
| Han | 0.982 (0.018) | 0.984 (0.015) | 0.982 (0.018) | 0.984 (0.015) | 0.00 |
| Male | 0.784 (0.0170 | 0.574 (0.245) | 0.784 (0.170) | 0.574 (0.245) | 0.00 |
| Married | 0.918 (0.074) | 0.859 (0.121) | 0.919 (0.074) | 0.859 (0.121) | 0.13 |
| Religious | 0.023 (0.023) | 0.043 (0.041) | 0.023 (0.023) | 0.043 (0.041) | 0.24 |
| Elementary | 0.077 (0.068) | 0.140 (0.120) | 0.073 (0.068) | 0.140 (0.120) | 0.00 |
| Junior High | 0.264 (0.194) | 0.382 (0.236) | 0.262 (0.194) | 0.382 (0.236) | -0.00 |
| Senior High | 0.190 (0.154) | 0.222 (0.173) | 0.190 (0.154) | 0.222 (0.173) | -0.00 |
| Technical School | 0.015 (0.015) | 0.033 (0.032 | 0.016 (0.015) | 0.033 (0.032) | -0.00 |
| Vocational School | 0.125 (0.109) | 0.063 (0.060) | 0.125 (0.109) | 0.064 (0.060) | -0.00 |
| Three-year College | 0.192 (0.157) | 0.064 (0.062) | 0.195 (0.157) | 0.067 (0.062) | 0.00 |
| Formal College | 0.115 (0.103) | 0.040 (0.038) | 0.117 (0.103) | 0.040 (0.038) | -0.00 |
| Graduate | 0.003 (0.001) | 0.001 (0.003) | 0.003 (0.03) | 0.001 (0.001) | -0.00 |
| Observations | 390 | 1706 | 385 | 1,609 |  |

Notes: Variances in parentheses. (a) t-tests for equality of means in the two samples (before and after matching if option both is specified). T-tests are based on a regression of the variable on a treatment indicator. Before matching or on raw samples this is an unweighted regression on the whole sample, after matching the regression is weighted using the matching weight variable _weight or user-given weight variable in mweight and based on the on-support sample. T-tests are based on Rosenbaum and Rubin (1985).

**TABLE A.1-3** Balancing Post Matching (CHIP 2002)

|  | Pre-matched Means (Variance) | | Post-matched Means (Variance) | | t-value for matched sample |
|---|---|---|---|---|---|
|  | Treatment | Control | Treatment | Control |  |
| Han | 0.949 (0.048) | 0.933 (0.063) | 0.949 (0.048) | 0.933 (0.063) | -0.18 |
| Male | 0.808 (0.155) | 0.805 (0.157) | 0.808 (0.155) | 0.805 (0.157) | -0.11 |
| Married | 0.963 (0.036) | 0.949 (0.048) | 0.963 (0.036) | 0.949 (0.048) | -0.13 |
| Primary | 0.061 (0.057) | 0.152 (0.129) | 0.061 (0.057) | 0.152 (0.129) | -0.00 |
| Middle school | 0.269 (0.197) | 0.430 (0.245) | 0.269 (0.197) | 0.430 (0.245) | -0.03 |
| High School | 0.219 (0.171) | 0.217 (0.170) | 0.219 (0.171) | 0.217 (0.170) | -0.00 |
| Technical School | 0.124 (0.109) | 0.067 (0.062) | 0.124 (0.109) | 0.067 (0.062) | -0.11 |
| Vocational School | 0.204 (0.162) | 0.066 (0.062) | 0.204 (0.162) | 0.066 (0.062) | 0.03 |
| College | 0.109 (0.097) | 0.026 (0.025) | 0.109 (0.097) | 0.026 (0.025) | 0.11 |
| Urban | 0.703 (0.209) | 0.503 (0.250) | 0.704 (0.209) | 0.506 (0.250) | -0.05 |
| Observations | 3,600 | 8,217 | 3,600 | 8,217 |  |

Notes: Variances in parentheses. (a) t-tests for equality of means in the two samples (before and after matching if option both is specified). T-tests are based on a regression of the variable on a treatment indicator. Before matching or on raw samples this is an unweighted regression on the whole sample, after matching the regression is weighted using the matching weight variable _weight or user-given weight variable in mweight and based on the on-support sample. Urbanicity was dropped in the matching procedure for CHIP 2002 since it was perfectly correlated with the occupational binary variables. T-tests are based on Rosenbaum and Rubin (1985).

**TABLE A.1-4** Balancing Post Matching (CGSS 2003)

|  | Pre-matched Means (Variance) | | Post-matched Means (Variance) | | $t$-value for matched sample |
|---|---|---|---|---|---|
|  | Treatment | Control | Treatment | Control |  |
| Han | 0.941 (0.056) | 0.945 (0.052) | 0.941 (0.056) | 0.945 (0.052) | -0.00 |
| Male | 0.463 (0.250) | 0.541 (0.248) | 0.463 (0.250) | 0.541 (0.248) | -0.00 |
| Married | 0.952 (0.046) | 0.843 (0.133) | 0.952 (0.046) | 0.843 (0.133) | -0.00 |
| Primary | 0.066 (0.062) | 0.132 (0.115) | 0.066 (0.062) | 0.132 (0.115) | 0.00 |
| Middle school | 0.131 (0.115) | 0.317 (0.217) | 0.131 (0.115) | 0.317 (0.217) | 0.00 |
| Highschool | 0.163 (0.137) | 0.189 (0.153) | 0.163 (0.137 | 0.189 (0.153) | -0.00 |
| Vocational | 0.007 (0.007) | 0.033 (0.032) | 0.007 (0.007) | 0.033 (0.032) | 0.00 |
| Technical | 0.166 (0.139) | 0.095 (0.086) | 0.166 (0.139) | 0.095 (0.086) | 0.00 |
| Junior college | 0.253 (0.189) | 0.135 (0.117) | 0.253 (0.189) | 0.135 (0.117) | -0.00 |
| College | 0.193 (0.157) | 0.059 (0.056) | 0.193 (0.157) | 0.059 (0.056) | -0.00 |
| Grad | 0.007 (0.007) | 0.005 (0.005) | 0.007 (0.007) | 0.005 (0.05) | 0.00 |
| Observations | 289 | 4,202 | 289 | 4,202 |  |

Notes: Variances in parentheses. (a) t-tests for equality of means in the two samples (before and after matching if option both is specified). T-tests are based on a regression of the variable on a treatment indicator. Before matching or on raw samples this is an unweighted regression on the whole sample, after matching the regression is weighted using the matching weight variable _weight or user-given weight variable in mweight and based on the on-support sample. T-tests are based on Rosenbaum and Rubin (1985).

**TABLE A.1-5** Balancing Post Matching (CGSS 2013)

|  | Pre-matched Means (Variance) | | Post-matched Means (Variance) | | $t$-value for matched sample |
|---|---|---|---|---|---|
|  | Treatment | Control | Treatment | Control |  |
| Han | 0.921 (0.073) | 0.912 (0.080) | 0.921 (0.073) | 0.912 (0.080) | -0.08 |
| Male | 0.747 (0.189) | 0.519 (0.250) | 0.747 (0.189) | 0.519 (0.250) | -0.05 |
| Married | 0.833 (0.139) | 0.782 (0.170) | 0.833 (0.139) | 0.782 (0.170) | -0.00 |
| Primary | 0.092 (0.084) | 0.233 (0.179) | 0.092 (0.084) | 0.233 (0.179) | 0.00 |
| Middle school | 0.197 (0.158) | 0.314 (0.215) | 0.197 (0.158) | 0.314 (0.215) | 0.00 |
| High School | 0.131 (0.114) | 0.113 (0.100) | 0.131 (0.114) | 0.113 (0.100) | 0.00 |
| Vocational School | 0.088 (0.080) | 0.048 (0.046) | 0.088 (0.080) | 0.048 (0.046) | 0.00 |
| Technical School | 0.016 (0.016) | 0.024 (0.023) | 0.016 (0.016) | 0.024 (0.023) | -0.00 |
| Junior College | 0.191 (0.154) | 0.072 (0.067) | 0.191 (0.154) | 0.072 (0.067) | -0.06 |
| College | 0.217 (0.170) | 0.053 (0.050) | 0.217 (0.170) | 0.053 (0.050) | -0.00 |
| Graduate School | 0.024 (0.023) | 0.005 (0.005) | 0.024 (0.023) | 0.005 (0.005) | 0.14 |
| Observations | 1,060 | 8,011 | 1,060 | 8,011 |  |

Notes: Variances in parentheses. (a) t-tests for equality of means in the two samples (before and after matching if option both is specified). T-tests are based on a regression of the variable on a treatment indicator. Before matching or on raw samples this is an unweighted regression on the whole sample, after matching the regression is weighted using the matching weight variable _weight or user-given weight variable in mweight and based on the on-support sample. T-tests are based on Rosenbaum and Rubin (1985).

**TABLE A.2** Over-identification Test for Covariate Balancing

| Survey Source /Year | CHIP 1988[a] | CHS 1993[b] | CHIP 2002[c] | CGSS 2003[d] | CGSS 2013[e] |
|---|---|---|---|---|---|
| | (1) | (2) | (3) | (4) | (5) |
| chi2(12)= 16.0546 | 60.4384 | 16.0546 | 46.053 | 22.631 | 56.39 |
| Prob > chi2 | 0.00 | 0.1887 | 0.0000 | 0.031 | 0.000 |
| Observations | 19,323 | 1,994 | 11,817 | 4,491 | 9,071 |

Notes: A formal test based on Imai and Ratkovic (2014) tests the null hypothesis that the IPW model balanced the covariates used in matching. ***, ** and * indicate significance at 1, 5 and 10%, respectively.

**TABLE A.3** Earnings Equation (2SLS)

| Dependent Variable: | ln (Monthly Earnings, in RMBs) | |
|---|---|---|
| Survey Source /Year: | CHIP 2002[a,b] | CGSS 2013[a] |
| | (1) | (2) |
| Communist Party | 0.170 | 0.277 |
| Membership | (0.388) | (0.465) |
| District Fixed Effects | YES | YES |
| F-statistic | 66.87 | 345.07 |
| R-squared | 0.060 | 0.3489 |
| Observations | 6,706 | 9,071 |

Notes: (a) The instrumental variable is parental Communist Party affiliation. (b) Based on the urban sample. ***, ** and * indicate significance at 1, 5 and 10%, respectively.

# Online Appendix B

TABLE B.1 Propensity Stratified Regressions

| Propensity Score Strata | Coefficient Estimate | | | | |
|---|---|---|---|---|---|
| | CHIP 1988 | CHS 1988 | CHIP 2002 | CGSS 2003 | CGSS 2013 |
| | (1) | (2) | (3) | (4) | (5) |
| 1 | 0.230 | 0.061 | -0.088 | 0.345*** | 0.545*** |
| | (0.272) | (0.089) | (0.631) | (0.095) | (0.124) |
| 2 | 0.168 | 0.107 | 0.343*** | 0.247** | 0.254** |
| | (0.141) | (0.083) | (0.106) | (0.112) | (0.098) |
| 3 | 0.232*** | 0.132*** | 0.323*** | 0.274*** | 0.075 |
| | (0.037) | (0.047) | (0.056) | (0.083) | (0.108) |
| 4 | 0.447* | 0.087 | 0.199 | 0.219*** | -0.066 |
| | (0.263) | (0.062) | (0.155) | (0.052) | (0.073) |
| 5 | 0.152*** | 0.033 | 0.172*** | 0.163 | 0.049 |
| | (0.053) | (0.082) | (0.035) | (0.163) | (0.071) |
| 6 | -0.080 | 0.150*** | 0.160* | 0.194** | 0.034 |
| | (0.200) | (0.078) | (0.091) | (0.065) | (0.073) |
| 7 | 0.177 | 0.007 | 0.031 | -0.313*** | -0.214** |
| | (0.350) | (0.052) | (0.035) | (0.095) | (0.079) |
| | | | | | |
| Observations | 51,681 | 1,994 | 11,887 | 4,491 | 9,071 |

Notes: Standard Errors in parantheses. ***, ** and * indicate significance at 1, 5 and 10%, respectively.

**TABLE B.2 Matching Algorithms**

| | Dependent Variable: Monthly Earnings (in RMB) | | | | |
|---|---|---|---|---|---|
| Algorithm Method | NN 2:1 Matching with replacement | NN 1:1 Matching Mahalonobis | IPW | Radius Caliper (0.20) | Kernel |
| | (1) | (2) | (3) | (4) | (5) |
| **Panel A (CHIP 1988):** | | | | | |
| Communist Party Membership | 0.177*** | 0.181*** | 0.189*** | 0.191*** | 0.189*** |
| | (0.010) | (0.010) | (0.009) | (0.009) | (0.009) |
| Controls | YES | YES | YES | YES | YES |
| Observations | 19,323 | 19,323 | 19,323 | 19,323 | 19,323 |
| **Panel B (CHS 1993):** | | | | | |
| Communist Party Membership | 0.094*** | 0.090*** | 0.085*** | 0.131*** | 0.0825*** |
| | (0.024) | (0.024) | (0.024) | (0.024) | (0.025) |
| Controls | YES | YES | YES | YES | YES |
| Observations | 1,994 | 1,994 | 1,994 | 1,994 | 1,994 |
| **Panel C (CHIP 2002):** | | | | | |
| Communist Party Membership | 0.188*** | 0.191*** | 0.209*** | 0.188*** | 0.209*** |
| | (0.022) | (0.023) | (0.022) | (0.022) | (0.022) |
| Controls | YES | YES | YES | YES | YES |
| Observations | 11,817 | 11,817 | 11,817 | 11,817 | 11,817 |
| **Panel D (CGSS 2003):** | | | | | |
| Communist Party Membership | 0.233*** | 0.237*** | 0.245*** | 0.237*** | 0.227*** |
| | (0.032) | (0.032) | (0.031) | (0.032) | (0.035) |
| Controls | YES | YES | YES | YES | YES |
| Observations | 4,491 | 4,491 | 4,491 | 4,491 | 4,491 |
| **Panel E (CGSS 2013):** | | | | | |
| Communist Party Membership | 0.212*** | 0.214*** | 0.253*** | 0.199*** | 0.257*** |
| | (0.052) | (0.052) | (0.047) | (0.054) | (0.048) |
| Controls | YES | YES | YES | YES | YES |
| Observations | 9,071 | 9,071 | 9,071 | 9,071 | 9,071 |

Notes: Variables used in estimating propensity score: sex dummy (1 if respondent is male and 0 if female), ethnicity dummy (1 if respondent is of the Han majority ethnicity and 0 if of a minority ethnicity), married dummy (1 if the respondent is married and 0 if single), dummies for education (1 if the respondent achieved the specified level of education and 0 if not) and religion (1 if the respondent is religious and 0 if town). Values can be interpreted as the percent change in monthly earnings. Communist party membership in a dummy equal to 1 if the subject is a member of the Communist party and 0 if otherwise. Standard Errors in Parentheses.
***, ** and * indicate significance at 1, 5 and 10%, respectively.

**TABLE B.3** Quantile Regressions

| | Dependent Variable: Monthly Earnings (in RMB) | | |
|---|---|---|---|
| | 0.25 Quantile | 0.50 Quantile | 0.75 Quantile |
| | (1) | (2) | (3) |
| Panel A (CHIP 1988): | | | |
| Communist Party Membership | 0.194*** | 0.159*** | 0.135*** |
| | (0.008) | (0.007) | (0.008) |
| Controls | YES | YES | YES |
| Observations | 19,323 | 19,323 | 19,323 |
| Panel B (CHS 1993): | | | |
| Communist Party Membership | 0.102*** | 0.040 | 0.029 |
| | (0.315) | (0.030) | (0.031) |
| Controls | YES | YES | YES |
| Observations | 1,994 | 1,994 | 1,994 |
| Panel C (CHIP 2002): | | | |
| Communist Party Membership | 0.136*** | 0.115*** | 0.119*** |
| | (0.030) | (0.022) | (0.020) |
| Controls | YES | YES | YES |
| Observations | 11,887 | 11,887 | 11,887 |
| Panel D (CGSS 2003): | | | |
| Communist Party Membership | 0.038*** | 0.020 | 0.033 |
| | (0.056) | (0.036) | (0.038) |
| Controls | YES | YES | YES |
| Observations | 9,071 | 9,071 | 9,071 |
| Panel E (CGSS 2013): | | | |
| Communist Party Membership | 0.223*** | 0.233*** | 0.182*** |
| | (0.068) | (0.051) | (0.053) |
| Controls | YES | YES | YES |
| Observations | 4,491 | 4,491 | 4,491 |

Notes: ***, ** and * indicate significance at 1, 5 and 10%, respectively.

**TABLE B.4-1** Rosenbaum Bounds (CHIP 1988)

| Gamma | Sig+ | Sig- | t-hat+ | t-hat- | CI+ | CI- |
|---|---|---|---|---|---|---|
| 1 | 0 | 0 | 0.165212 | 0.165212 | 0.157635 | 0.172853 |
| 2 | 0 | 0 | 0.088251 | 0.243723 | 0.080331 | 0.251972 |
| 3 | 0 | 0 | 0.044552 | 0.289714 | 0.036123 | 0.298698 |
| 4 | 0.001225 | 0 | 0.014213 | 0.322239 | 0.00511 | 0.332039 |
| 5 | 0.973904 | 0 | -0.00905 | 0.347436 | -0.01875 | 0.357977 |
| 6 | 1 | 0 | -0.02783 | 0.368033 | -0.03817 | 0.379353 |
| 7 | 1 | 0 | -0.04379 | 0.385449 | -0.05473 | 0.397537 |
| 8 | 1 | 0 | -0.05752 | 0.400675 | -0.06909 | 0.413432 |
| 9 | 1 | 0 | -0.06966 | 0.414064 | -0.08193 | 0.427514 |
| 10 | 1 | 0 | -0.08066 | 0.426081 | -0.09346 | 0.440351 |

Notes: gamma - log odds of differential assignment due to unobserved factors. sig+ - upper bound significance level. sig- - lower bound significance level. t-hat+ - upper bound Hodges-Lehmann point estimate. t-hat- - lower bound Hodges-Lehmann point estimate. CI+ - upper bound confidence interval (a= .95). CI- - lower bound confidence interval (a= .95)

**TABLE B.4-2** Rosenbaum Bounds (CHS 1993)

| Gamma | Sig+ | Sig- | t-hat+ | t-hat- | CI+ | CI- |
|---|---|---|---|---|---|---|
| 1 | 4.30E-10 | 4.30E-10 | 0.121024 | 0.121024 | 0.084781 | 0.157868 |
| 2 | 0.310705 | 0 | 0.01046 | 0.234923 | -0.03038 | 0.276998 |
| 3 | 0.996879 | 0 | -0.0516 | 0.299525 | -0.0908 | 0.347857 |
| 4 | 1 | 0 | -0.09083 | 0.347921 | -0.13138 | 0.400047 |
| 5 | 1 | 0 | -0.11937 | 0.38523 | -0.16277 | 0.434136 |
| 6 | 1 | 0 | -0.1428 | 0.411439 | -0.1875 | 0.464085 |
| 7 | 1 | 0 | -0.1618 | 0.433158 | -0.2081 | 0.491797 |
| 8 | 1 | 0 | -0.17666 | 0.451299 | -0.22549 | 0.51259 |
| 9 | 1 | 0 | -0.18966 | 0.469065 | -0.23943 | 0.532992 |
| 10 | 1 | 0 | -0.20179 | 0.484703 | -0.25231 | 0.547123 |

Notes: gamma - log odds of differential assignment due to unobserved factors. sig+ - upper bound significance level. sig- - lower bound significance level. t-hat+ - upper bound Hodges-Lehmann point estimate. t-hat- - lower bound Hodges-Lehmann point estimate. CI+ - upper bound confidence interval (a= .95). CI- - lower bound confidence interval (a= .95)

**TABLE B.4-3** Rosenbaum Bounds (CHIP 2002)

| Gamma | Sig+ | Sig- | t-hat+ | t-hat- | CI+ | CI- |
|---|---|---|---|---|---|---|
| 1 | 0 | 0 | 0.362942 | 0.362942 | 0.29483 | 0.43134 |
| 2 | 0.712764 | 0 | -0.02075 | 0.718146 | -0.0986 | 0.785225 |
| 3 | 1 | 0 | -0.25135 | 0.91118 | -0.33769 | 0.97952 |
| 4 | 1.00E+00 | 0 | -0.41325 | 1.03916 | -0.50755 | 1.10976 |
| 5 | 1 | 0 | -0.5407 | 1.13236 | -0.64361 | 1.20491 |
| 6 | 1 | 0 | -0.64415 | 1.20532 | -0.75438 | 1.27913 |
| 7 | 1 | 0 | -0.72989 | 1.26395 | -0.84877 | 1.33872 |
| 8 | 1 | 0 | -0.80675 | 1.31265 | -0.927 | 1.38968 |
| 9 | 1 | 0 | -0.87374 | 1.35433 | -0.99846 | 1.43385 |
| 10 | 1 | 0 | -0.92896 | 1.3905 | -1.0613 | 1.47297 |

Notes: gamma - log odds of differential assignment due to unobserved factors. sig+ - upper bound significance level. sig- - lower bound significance level. t-hat+ - upper bound Hodges-Lehmann point estimate. t-hat- - lower bound Hodges-Lehmann point estimate. CI+ - upper bound confidence interval (a= .95). CI- - lower bound confidence interval (a= .95)

**TABLE B.4-4** Rosenbaum Bounds (CGSS 2003)

| Gamma | Sig+ | Sig- | t-hat+ | t-hat- | CI+ | CI- |
|---|---|---|---|---|---|---|
| 1 | 0 | 0 | 0.480354 | 0.480354 | 0.395073 | 0.558735 |
| 2 | 4.00E-07 | 0 | 0.265161 | 0.679859 | 0.169027 | 0.762688 |
| 3 | 5.63E-03 | 0 | 0.135897 | 0.794804 | 0.030456 | 0.875794 |
| 4 | 0.183534 | 0 | 0.045364 | 0.866592 | -0.074153 | 0.959258 |
| 5 | 0.636796 | 0 | -0.022168 | 0.919445 | -0.152859 | 1.01638 |
| 6 | 0.915482 | 0 | -0.080858 | 0.966481 | -0.224391 | 1.06903 |
| 7 | 0.987802 | 0 | -0.133331 | 0.999042 | -0.28097 | 1.103 |
| 8 | 0.998736 | 0 | -0.173593 | 1.03078 | -0.334802 | 1.14278 |
| 9 | 0.999896 | 0 | -0.214004 | 1.063 | -0.384472 | 1.17452 |
| 10 | 0.999993 | 0 | -0.238746 | 1.0828 | -0.426642 | 1.1951 |

Notes: gamma - log odds of differential assignment due to unobserved factors. sig+ - upper bound significance level. sig- - lower bound significance level. t-hat+ - upper bound Hodges-Lehmann point estimate. t-hat- - lower bound Hodges-Lehmann point estimate. CI+ - upper bound confidence interval (a= .95). CI- - lower bound confidence interval (a= .95)

**TABLE B.4-5** Rosenbaum Bounds (CGSS 2013)

| Gamma | Sig+ | Sig- | t-hat+ | t-hat- | CI+ | CI- |
|---|---|---|---|---|---|---|
| 1 | 0 | 0 | 0.422942 | 0.422942 | 0.370801 | 0.468287 |
| 2 | 6.50E-08 | 0 | 0.169048 | 0.65102 | 0.111045 | 0.699623 |
| 3 | 3.74E-01 | 0 | 0.012907 | 0.776261 | -0.05407 | 0.825773 |
| 4 | 9.99E-01 | 0 | -0.10476 | 0.862256 | -0.18496 | 0.915829 |
| 5 | 1 | 0 | -0.20274 | 0.926319 | -0.28698 | 0.983245 |
| 6 | 1 | 0 | -0.27867 | 0.974446 | -0.37912 | 1.03708 |
| 7 | 1 | 0 | -0.35319 | 1.02181 | -0.4511 | 1.08279 |
| 8 | 1 | 0 | -0.41144 | 1.05917 | -0.52012 | 1.12308 |
| 9 | 1 | 0 | -0.46545 | 1.08799 | -0.58164 | 1.15766 |
| 10 | 1 | 0 | -0.50778 | 1.11764 | -0.63061 | 1.19141 |

Notes: gamma - log odds of differential assignment due to unobserved factors. sig+ - upper bound significance level. sig- - lower bound significance level. t-hat+ - upper bound Hodges-Lehmann point estimate. t-hat- - lower bound Hodges-Lehmann point estimate. CI+ - upper bound confidence interval (a= .95). CI- - lower bound confidence interval (a= .95)